%% file: CFF_ARMA.tex
\documentclass[12pt,a4paper]{article}
\usepackage[DIV=13]{typearea}
\usepackage[onehalfspacing]{setspace}
\input{Definitions}

\title{Approximate State Space Modelling of Unobserved Fractional Components}
\author[1,2]{Tobias Hartl}
\author[3]{Roland Weigand\footnote{Corresponding author. E-Mail: roland.weigand@posteo.de}}
\affil[1]{University of Regensburg, 93053 Regensburg, Germany}
\affil[2]{Institute for Employment Research (IAB), 90478 Nuremberg, Germany}
\affil[3]{AOK Bayern, 93055 Regensburg, Germany}
\date{March 2020}

\begin{document}
\input{Sec0_Abstract.tex}

\input{Sec1_Introduction.tex}

\input{Sec2_SSF.tex}

\input{Sec3_ML.tex}

\input{Sec4_MC.tex}
\input{Sec5_Concl.tex}
\newpage
\begin{spacing}{1.2}
\bibliographystyle{dcu}
\bibliography{../Literatur/literatur.bib}
\end{spacing}
\newpage
\input{SecA1_App.tex}
\clearpage
\renewcommand{\arraystretch}{1.3}
\clearpage
\input{SecA2_Tables_new.tex}

\clearpage
\input{SecA3_Figures.tex}

\end{document}

%% file: Definitions.tex
\usepackage[english]{babel}
\usepackage[ansinew]{inputenc}
\usepackage{amssymb,amsmath,bm,array,dsfont,graphicx,natbib}
\usepackage[colorlinks,citecolor=black,linkcolor=black,urlcolor=black]{hyperref}
\usepackage[hang,tight]{subfigure}
\usepackage{authblk}

\newcommand{\newoperator}[3]{\newcommand*{#1}{\mathop{#2}#3}}
\newcommand{\renewoperator}[3]{\renewcommand*{#1}{\mathop{#2}#3}}
\newcommand{\mA}{A}

\newcommand{\mB}{B}

\newcommand{\mC}{C}

\newcommand{\mD}{D}
\newcommand{\vd}{d}
\newcommand{\mE}{E}

\newcommand{\mF}{F}
\newcommand{\vf}{f}
\newcommand{\mG}{G}
\newcommand{\vg}{g}
\newcommand{\mH}{H}

\newcommand{\mI}{I}

\newcommand{\mL}{L}

\newcommand{\mM}{M}

\newcommand{\mQ}{Q}

\newcommand{\mR}{R}

\newcommand{\mS}{S}

\newcommand{\mT}{T}

\newcommand{\vu}{u}
\newcommand{\mV}{V}
\newcommand{\vv}{v}

\newcommand{\vx}{x}

\newcommand{\vy}{y}
\newcommand{\mZ}{Z}

\newcommand{\valpha}{\alpha}

\newcommand{\vgamma}{\gamma}

\newcommand{\vepsi}{\epsi}

\newcommand{\veta}{\eta}
\newcommand{\vtheta}{\theta}

\newcommand{\vlambda}{\lambda}
\newcommand{\vmu}{\mu}

\newcommand{\vxi}{\xi}

\newcommand{\mGamma}{\varGamma}

\newcommand{\mLambda}{\varLambda}
\newcommand{\mXi}{\varXi}
\newcommand{\mPi}{\varPi}

\renewoperator{\Re}{\mathrm{Re}}{\nolimits}
\renewoperator{\Im}{\mathrm{Im}}{\nolimits}
\makeatletter
\newcommand{\rd}{\@ifnextchar^{\DIfF}{\DIfF^{}}}
\def\DIfF^#1{%
   \mathop{\mathrm{\mathstrut d}}%
   \nolimits^{#1}\gobblespace}
\def\gobblespace{\futurelet\diffarg\opspace}
\def\opspace{%
   \let\DiffSpace\!%
   \ifx\diffarg(%
   \let\DiffSpace\relax
   \else
   \ifx\diffarg[%
   \let\DiffSpace\relax
   \else
   \ifx\diffarg\{%
   \let\DiffSpace\relax
   \fi\fi\fi\DiffSpace}

\newcommand{\E}{\operatorname{E}}

\newcommand{\median}{\operatorname{median}}
\newcommand{\Var}{\operatorname{Var}}
\newcommand{\Corr}{\operatorname{Corr}}

\newcommand{\diag}{\operatorname{diag}}

\newoperator{\ip}{\mathrm{int}}{\nolimits}

\newcommand{\MSE}{\operatorname{MSE}}

\newcommand{\tr}{\operatorname{tr}}
\renewcommand{\vec}{\operatorname{vec}}

\newcommand{\epsi}{\varepsilon}

\newcommand{\vzeros}{0}
\newcommand{\mZeros}{0}
\newcommand{\mzeros}{\mZeros}

\newcommand{\beq}{\begin{equation}}
\newcommand{\eeq}{\end{equation}}
\newcommand{\bal}{\begin{align*}}
\newcommand{\eal}{\end{align*}}

\newcommand{\bmat}{\begin{bmatrix}}
\newcommand{\emat}{\end{bmatrix}}

\newcommand{\bsmat}{\begin{smallmatrix}}
\newcommand{\esmat}{\end{smallmatrix}}

%% file: Sec0_Abstract.tex
\maketitle

\thispagestyle{empty}
\setcounter{page}{0}

\paragraph{\bf Abstract.}

We propose convenient inferential methods for potentially nonstationary multivariate unobserved components models with fractional integration and cointegration. Based on finite-order ARMA approximations in the state space representation, maximum likelihood estimation can make use of the EM algorithm and related techniques. The approximation outperforms the frequently used autoregressive or moving average truncation, both in terms of computational costs and with respect to approximation quality. Monte Carlo simulations reveal good estimation properties of the proposed methods for processes of different complexity and dimension. 

\paragraph{\bf Keywords.}

Long memory, fractional cointegration, state space, unobserved components.

\paragraph{\bf JEL-Classification.}

C32, C51, C53, C58.

\newpage

%% file: Sec1_Introduction.tex
\section{Introduction}

Fractionally integrated time series models have gained significant interest in recent decades. In possibly nonstationary multivariate setups, which arguably bear most potential e.g.\ for assessing macroeconomic linkages, and which are essential for the joint modelling of financial processes, several parametric models have been explored. Among the most popular are the fractionally integrated VAR model \citep{Nie2004b}, the triangular fractional cointegration model of \citet{RobHua2003} and the cointegrated VAR$_{d,b}$ model of \citet{Joh2008}.

Meanwhile, also models with unobserved fractional components have proven useful, as empirical and methodological work by \citet{RayTsa2000}, \citet{Mor2004}, \citet{CheHur2006}, \citet{Mor2007} and \citet{LucVer2015} documents. 
The unobserved fractional components model allows for a generalization of the classic trend-cycle decomposition, where the long-run component is typically assumed to be I(1). As \citet{HarTscWeb2019} show, the model can be used to test the I(1) assumption against a fractional alternative. Furthermore, unobserved fractional components allow the formulation of parsimonious models, like factor models, in an interpretable way.

These methods offer a variety of potential applications to empirical researchers. Long-run components of GDP, (un-)employment, and inflation are typically estimated via unobserved components models that restrict the integration order to unity \citep[cf.\ e.g.][]{MorNelZi2003, DomGom2006, KliWeb2016}. 
Since there is comprehensive evidence for long memory in these variables \citep[see e.g.][]{HasWol1995, DijFraPa2002, TscWebWe2013} unobserved fractional components may provide new insights regarding the form and persistence of the long-run components. On the other hand, fractionally integrated factor models are constructed straightforwardly using fractional unobserved components. They can be used to assess fractional cointegration relations and for forecasting, as \citet{HarWei2018a} demonstrate.

Inferential methods for such unobserved fractional components are the subject of this paper. So far, the bulk of empirical work in this field has been conducted in a semiparametric setting, which may be explained by the high computational and implementation cost of state-of-the-art parametric approaches such as simulated maximum likelihood \citep{MesKooOo2016}. Especially for models of relatively high dimensions or with a rich dynamic structure, there is a lack of feasible estimation methods. Furthermore, in most empirical applications, methods are required to smoothly handle nonstationary cases alongside stationary ones.

We consider a computationally straightforward parametric treatment of fractional unobserved components models in state space form. An approximation of potentially nonstationary fractionally integrated series using finite-order ARMA structures is suggested. This procedure outperforms the more commonly used truncation of fractional processes \citep[cf.][]{ChaPal1998} by providing a substantial reduction of the state dimension and hence of computational costs for a desired approximation quality. We derive both, the log likelihood and an analytical expression for the corresponding score. Hence, parameter estimation by means of the EM algorithm and gradient-based optimization make the approach feasible even for high dimensional datasets. In Monte Carlo simulations we study the performance of the proposed methods and quantify the accuracy of our state space approximation. For fractionally integrated and cointegrated processes of different dimensions, we find promising finite-sample estimation properties also in comparison to alternative techniques, namely the exact local Whittle estimator, narrow band least squares and exact state space methods. By using a parameter-driven state space approach, our setup inherits several additional favorable properties: Missing values are treated seamlessly, several types of structural time series components such as trends, seasons and noise can be added without effort, and a wide variety of possibly nonlinear or non-Gaussian observation schemes may be straightforwardly implemented; see \citet{Har1991,DurKoo2012}.

In this paper we apply the proposed approximation scheme to a $p$-dimensional observed time series $\vy_t$, which is driven by a fractional components (FC) process as defined by \citet{HarWei2018a},
\beq \label{eq:FC}
\vy_t = \mLambda \vx_t + \vu_t, \qquad t=1,\ldots,n.
\eeq
Here, $\mLambda$ is a $p \times s$ coefficient matrix with full column rank, the latent process $x_t = (x_{1t}, ..., x_{st})'$ holds the purely fractional components which are driven by a noise process $\xi_{t}=(\xi_{1t}, ..., \xi_{st})'\sim \mathrm{NID}(0, \mI)$,
and $\vu_t$ holds the short memory components. 

More precisely, while the stationary series $\vu_t$ is only required to have a finite state space representation, the components of the $s$-dimensional $\vx_t$ are fractionally integrated noise according to
\beq \label{eq:FC_frac}
\Delta^{d_j} x_{jt} = \xi_{jt}, \qquad j=1,\ldots,s, 
\eeq
where for a generic scalar $d$, the fractional difference operator is defined by
\beq \label{eq:fracdiff}
\Delta^d = (1-L)^d = \sum_{j=0}^{\infty} \pi_j(d)L^j, \quad \pi_0(d) =1, \quad \pi_j(d) =
\frac{j-1-d}{j}  \pi_{j-1}(d), \; j\geq 1,
\eeq
and $L$ denotes the lag or backshift operator, $Lx_t = x_{t-1}$. We adapt a nonstationary type II solution of these processes \citep{Rob2005} and hence treat $d_j\geq0.5$ alongside the asymptotically stationary case $d_j<0.5$ in a continuous setup.

The fractional unobserved components framework captures univariate and multivariate processes with both long-run and short-run dynamics, fractional cointegration and polynomial cointegration, as well as possibly high-dimensional processes with factor structure. It allows for an intuitive additive separation of long run and short run components, i.e. cyclical and trend components in business cycle analysis, while obtaining similar flexibility as (cointegrated) multivariate ARFIMA models. See \citet{HarWei2018a} for the relation of the FC model to several other fractional integration setups.

The paper is organized as follows. Section \ref{sec:estimation} discusses the state space form, while section \ref{sec:ml} outlines maximum likelihood estimation. In section \ref{sec:mc}, the estimation properties are investigated by means of Monte Carlo experiments before section \ref{sec:conclusion} concludes.

%% file: Sec2_SSF.tex
\section{The approximate state space form} \label{sec:estimation}


\subsection{Approximating nonstationary fractional integration} \label{sec:arma}

Unlike the stationary long-memory processes considered in the literature, e.g., by \citet{ChaPal1998}, \citet{HsuRayBr1998}, \citet{HsuBre2003}, \citet{Bro2007}, \citet{MesKooOo2016} as well as \citet{GraMag2012}, our nonstationary type II specification of fractional integration is straightforwardly represented in its exact state space form by setting starting values of the latent fractional process to zero, $x_{jt}=0$ for $t\leq 0$. The solution for $x_{jt}$ is based on the truncated operator $\Delta_+^{-d_j}$ \citep{Joh2008} and given by
\[
x_{jt} = \Delta_+^{-d_j} \xi_{jt} = \sum_{i=0}^{t-1} \pi_i(-d_j) \xi_{j,t-i}, \qquad j=1,\ldots,s.
\]
For a given sample size $n$, $\vx_t$ has an autoregressive structure with coefficient matrices $\mPi_j^{d}$  $= \diag(\pi_j(d_1), \ldots, \pi_j(d_s))$, $j=1,\ldots,n$. Thus, a Markovian state vector embodying $\vx_{t}$ has to include $n-1$ lags of $\vx_{t}$ and is initialized deterministically with $\vx_{-n+1}=\ldots=\vx_{0}=0$. In principle, this exact state space form can be used to compute the Kalman filter, to evaluate the likelihood and to estimate the unknown model parameters by nonlinear optimization routines. Since the state vector is at least of dimension $s\cdot n$, this can become computationally very costly, particularly in large samples and for a large number $s$ of fractional components, which makes a treatment of the system in its exact state space representation practically infeasible for a wide range of relevant applications.

For non-negative integration orders, note that the $x_t$ can be generalized to have non-zero starting values $x_0$. In that case, $x_t$ is initialized via a diffuse initial state vector. Details on the initialization of nonstationary components are given in \citet[][]{Koo1997}. Deterministic components in $x_t$ are handled straightforwardly by defining $x_{jt}^* := x_{jt} + \mu_{jt}$ where $\Var(\mu_{jt})=0$ $\forall j = 1,..., s$. Depending on the integration order $d_j$ the contribution of $\mu_{jt}$ on $\vy_t$ converges to a trend of degree $\lfloor d_j \rfloor$ as $t \rightarrow \infty$.

The literature on stationary long-memory processes has considered approximations based on a truncation of the autoregressive representation, considering only $m$ lags of $\vx_t$ for $m<n$ in the transition equation (i.e., setting all autoregressive coefficients to zero for $j>m$). Alternatively, the moving average representation has been truncated to arrive at a feasible state space model; see \citet{Pal2007}, sections 4.2 and 4.3.

Instead, we will apply ARMA approximations to the fractional state vectors, which provide a better approximation quality than the autoregressive or moving average truncation. An ARMA approximation of long-memory processes has been considered in the importance sampling frameworks of \citet{HsuBre2003} and \citet{MesKooOo2016}, but, arguably due to their computational burdens, did not find usage in applied research so far. In our setup, where fractional integration appears in the form of purely fractional components rather than ARFIMA processes, this approach is particularly convenient. In contrast to recent attempts to approximate ARFIMA processes by ARMA ones \citep[discussed e.g.\ by][]{BasChaPa2001}, we do not freely estimate all ARMA parameters but only $d$, and thus retain the original parsimonious parameterization of the process.

As a (nonstationary) approximation of a generic univariate $x_t = \Delta_+^{-d}\xi_t$, we consider the process
\beq \label{eq:arma_1}
\tilde{x}_t = \left[\frac{(1+ m_1L + \ldots + m_wL^w)}{(1 - a_1L - \ldots - a_{v}L^v)}\right]_+\xi_t = \sum_{j=0}^{n-1} \tilde{\psi}_j(\varphi) \xi_{t-j},
\eeq
for finite $v$ and $w$, where $\varphi:=(a_1,\ldots,a_v,m_1,\ldots,m_w)'$ and all $a_i$ and $m_j$ are made functionally dependent on $d$ to approximate $x_t$ by $\tilde{x}_t$. In order to determine the parameters $\varphi$, we minimize the distance between $x_t$ and $\tilde x_t$, using the mean squared error (MSE) over $t=1,\ldots,n$ as the distance measure. For given $t$, $d$ and $\varphi$, we observe
\beq
\tilde{x}_t -x_t = \sum_{j=0}^{t-1} \tilde{\psi}_j(\varphi) \xi_{t-j} - \sum_{j=0}^{t-1} \psi_j(d) \xi_{t-j} = \sum_{j=0}^{t-1} (\tilde{\psi}_j(\varphi)-\psi_j(d)) \xi_{t-j}. \label{approximation_error}
\eeq
Hence, the MSE for period $t$ is given by
\[
E[ (\tilde{x}_t -x_t)^2] = Var(\xi_t) \sum_{j=0}^{t-1} (\tilde{\psi}_j(\varphi)-\psi_j(d))^2,
\]
while averaging over all periods for a given sample size $n$ and ignoring the constant variance term yields the objective function for a given $d$,
\beq \label{eq:MSE_d}
\MSE_n^d(\varphi) = \frac{1}{n} \sum_{t=1}^n \sum_{j=0}^{t-1} (\tilde{\psi}_j(\varphi)-\psi_j(d))^2 = \frac{1}{n}\sum_{j=1}^{n} (n-j+1) (\tilde{\psi}_j(\varphi)-\psi_j(d))^2.
\eeq
The approximating ARMA coefficients are thus given by
\beq \label{eq:arma_2}
\hat{\varphi}_n(d) = \arg \min_{\varphi} \MSE_n^d(\varphi).
\eeq
To obtain the approximating ARMA coefficients in practice, we conduct the optimization \eqref{eq:arma_2} over a reasonable range of $d$, such as $d \in [-0.5; 2]$, for a given $n$. Computational details of the optimization are given in Appendix \ref{app:a1}. Interestingly, for $d < 1$, stationary ARMA coefficients provide the minimum MSE, while for $d\geq1$ we impose an appropriate number of unit roots to enhance the approximation quality.

To illustrate the results we plot the approximating ARMA(2,2) parameters as a function of $d$ for $n=500$; see figure \ref{fig:d2arma}. A closer look at the coefficients reveals that for $d>0$ typically both the autoregressive and the moving average polynomial have roots close to unity which nearly cancel out. For example, to approximate a process with $d=0.75$ we have $(1-1.932L +0.932L^2)\tilde{x}_t = (1-1.285L+ 0.306L^2) \xi_t$, which can be factorized as $(1- 0.999L) (1-0.933L)\tilde{x}_t = (1-0.970L)(1- 0.316L) \xi_t$. 
Despite their similarity, AR and MA roots do not cancel out for non-integer $d$, since the approximation quality is improved by additional free parameters. For integer integration orders the optimization yields $(1-0.953L)(1-0.477L)\tilde{x}_t = (1-0.953L)(1-0.477L)\xi_t$ for $d=0$ and $(1-L)(1-0.980L)\tilde{x}_t = (1-0.980L)(1+0.001L)\xi_t$ for $d=1$. Consequently, our ARMA-approximation is consistent with the finite representation of inter-integrated processes. 

To compare the ARMA($v$,$w$) approximations with $v=w \in \{1,2,3,4\}$ to a truncated AR($m$) process, we contrast the approximating impulse response function $\tilde{\psi}_j$ to the true one, $\psi_j(d)$, for a given $d$. The autoregressive truncation lag $m=50$ is used for our comparison, since this is among the largest values which we consider as feasible in a typical multivariate application. The result of this comparison is shown in figure \ref{fig:irarma} for $n=500$ and $d=0.75$. The autoregressive truncation approach gives the exact impulse responses for horizons $j\leq50$, but then tapers off too fast. The ARMA approximations improves significantly over the autoregressive truncation whenever $v=w\geq2$. For orders 3 or 4, the approximation error is even hardly visible. For the moving average truncation, the impulse responses equal zero for horizons exceeding the truncation lag (not shown).

To perform the comparison for different $d$, we plot the square root of the MSE \eqref{eq:MSE_d} as a function of $d$ for different approximation methods. For negative integration orders, as shown in figure \ref{fig:rmse1arma}, the moving average approach clearly outperforms the autoregression, while the ARMA method with orders $v=w>2$ are better. The moving average approximation becomes inaccurate, however, for the case $d>0$, and worse even than the autoregressive method as can be seen in figure \ref{fig:rmse2arma}. In contrast, the ARMA(3,3) and ARMA(4,4) approximations are well-suited to mimic fractional processes over the whole range of $d$. Further evidence in favor of the ARMA approximation will be presented in the Monte Carlo simulation of section \ref{ssec:mc1}.

\subsection{The state space representations} \label{sec:ssf}

Based on these methods we introduce the state space form of the multivariate model \eqref{eq:FC}, where each $x_{jt}$ is approximated by the ARMA approach. In the following we drop the tilde for the approximation of $x_{jt}$ for notational convenience. To cover the very general case, we allow for residual auto- and cross-correlation by modelling the latent $p$-dimensional short memory process $\vu_t$ via a stationary state space model, which can capture vector autoregressive, vector ARMA or factor models, among others, and include an additional noise term $\vepsi_t $. The model can be written in state space form as
\beq\label{eq:SSF}
\vy_t  = \mZ \valpha_t+ \vepsi_t, \qquad
\valpha_{t+1} = \mT \valpha_{t} + \mR \veta_t, \qquad \veta_t \sim \mathrm{NID}(0, \mQ), \qquad
\vepsi_t \sim \mathrm{NID}(0, \mH),
\eeq
where the states may be partitioned into $\valpha'_t =(\valpha^{(1)'}_t,\valpha^{(2)'}_t)$, the states related to the fractional and the stationary components, respectively.

Regarding the fractional part, we define $\mA_j^d:=\diag(\hat{a}_j(d_1),\ldots,\hat{a}_j(d_s))$ and $\mM_j^d:=\diag(\hat{m}_j(d_1),\ldots,\hat{m}_j(d_s))$ which contain the approximating AR and MA coefficients of the fractional noise introduced in section \ref{sec:arma}, while $\mA_{j}^d = \mzeros$ for $j>v$ and $\mM_{j}^d = \mzeros$ for $j>w$. Then $(\mI - \mA_1^d L -  ... - \mA_v^dL^v)(\mI + \mM_1^d L + ... + \mM_w^dL^w)^{-1}\tilde{x}_t = \vxi_t$. For a minimal state space representation, define $\vmu_t := (\mI + \mM_1^d L + ... + \mM_w^dL^w)^{-1}\tilde{x}_t$  such that $(\mI - \mA_1^d L -  ... - \mA_v^dL^v) \vmu_t = \vxi_t$. 
For $u=\max(v,w+1)$, the first part of the state vector is a $(us)$-dimensional process $\valpha^{(1)'}_t=(\vmu_t',\ldots, \vmu_{t-u+1}')$. Thus, $\valpha^{(1)}_{t+1} = \mT^{(1,1)} \valpha^{(1)}_{t} + \mR^{(1)} \veta_{t}^{(1)}$ with $\veta_{t}^{(1)} = \vxi_t$, $\mR^{(1)'} = (\mI, \mzeros, \ldots, \mzeros)^{'}$, $\mQ^{(1,1)} = \mI$ and
\[
\mT^{(1,1)}
=
\bmat
\mA_1^d& \mA_2^d &\ldots &\mA_u^d  \\
\mI &  &  &\mZeros   \\
\vdots & \ddots & &\vdots  \\
\mZeros & \ldots &\mI &\mZeros  \\
\emat.
\]
The observation equation for the fractional part is $\tilde{\vx}_{t} =  \vmu_t + \mM_1^d\vmu_{t-1}+ \ldots +\mM_{u-1}^d \vmu_{t-u}$, which enters the observed process $\vy_t$ 
through $\mLambda \tilde{\vx}_{t} = \mZ^{(1)}\valpha^{(1)}_t$. Thus, the observation matrix for the fractional part is
\[
\mZ^{(1)} =
\bmat
\mLambda & \mLambda \mM_1^d & \ldots & \mLambda \mM_{u-1}^d
\emat.
\]

For the nonfractional part, we allow a general specification with $\valpha^{(2)}_{t+1} = \mT^{(2,2)} \valpha^{(2)}_{t} + \mR^{(2)} \veta^{(2)}_t$ and $\vu_t = \mZ^{(2)} \valpha_{t}^{(2)}$, where the distribution of unknown parameters over $\mT^{(2,2)}$ and $\mZ^{(2)}$ reflect the choice of the specific model. Without loss of generality, we set $\mQ^{(2,2)} = \Var(\veta^{(2)}_t) = \mI$, so that scales and cross correlations of $\vu_t$ are determined by $\mZ^{(2)}$. The full state space model \eqref{eq:SSF} is given by an obvious definition of the system matrices as $\mZ = (\mZ^{(1)},\mZ^{(2)})$, $\mR' = (\mR^{(1)'},  \mR^{(2)'})$, $\mT = \diag( \mT^{(1,1)},  \mT^{(2,2)})$ and $\mQ = \mI$. The dynamics are complemented by the initial conditions for the states. From the definition of our type II fractional process we set fixed starting values such as $\valpha^{(1)}_0 =\vzeros$, while $\valpha^{(2)}_t$ is initialized by its stationary distribution.

The fractional components $\tilde{\vx}_t$ do not explicitly appear as states in this representation. However, filtered and smoothed states can be constructed using the relation $\tilde{\vx}_t = \vmu_{t}+\sum_{j=1}^{w} \mM^d_j\vmu_{t-j}$. To obtain conditional covariance matrices for $\tilde{\vx}_t$, it is more convenient to use an alternative state space form of the ARMA process, where the MA coefficients appear in $\mR^{(1,1)}$ rather than in $\mZ^{(1)}$; see \citet{DurKoo2012}, section 3.4. The current setup, however, is appropriate for estimating the parameters via the EM algorithm which is discussed in the next section.

%% file: Sec3_ML.tex
\section{Maximum likelihood estimation} \label{sec:ml}

The EM algorithm was proposed for maximum likelihood estimation of state space models by \citet{ShuSto1982} and \citet{WatEng1983}. Especially in the context of high-dimensional dynamic factor models with possibly more than hundred observable variables, i.e. $p>100$, this method has been found very useful in finding maxima of high-dimensional likelihood functions; see, e.g., \citet{QuaSar1993}, \citet{DozGiaRe2012} and \citet{JunKoo2014}. After rapidly locating an approximate optimum, the final steps until convergence are typically slow for the EM algorithm, and hence it has been suggested to switch to gradient-based methods with analytical expressions for the likelihood score at a certain step. 

We will present these algorithms for our fractional model and thereby extend existing treatments in the literature. For the model represented by \eqref{eq:SSF}, the matrices $\mT$ and $\mZ$ both nonlinearly depend on $\vd$ and other unknown parameters, so that there are nonlinear cross-equation restrictions linking the transition and the observation equation of the system.

The EM algorithm in general consists of two steps, which are repeated until convergence. In the E-step the expected complete data likelihood is computed, where the expectation is evaluated for a given set of parameters $\vtheta_{\{j\}}$, while the M-step maximizes this function to arrive at the parameters used in the next E-step, $\vtheta_{\{j+1\}}$. Thus, we define $Q(\vtheta,\tilde{\vtheta}) := \E_{\tilde{\vtheta}}\left[ l(\vtheta)\right]$, where in this section all expectation operators are understood as conditional on the data $\vy_1,\ldots,\vy_n$. In the course of the EM algorithm, after choosing suitable starting values $\vtheta_{\{1\}}$, the optimization $\vtheta_{\{j+1\}} = \arg \max_{\vtheta} Q(\vtheta, \vtheta_{\{j\}})$
is iterated for $j=1,2,\ldots$ until convergence.

To state the algorithm for the model defined by \eqref{eq:FC} and specified further in section \ref{sec:ssf}, we follow \citet{WuPaiHo1996} to obtain the expected complete data likelihood as
\begin{align} \label{eq:em_q}
	Q(\vtheta;\vtheta_{\{j\}}) = & -\frac{n}{2} \log|\mQ|  - \frac{1}{2} \tr \left[ \mR\mQ^{-1}\mR' (\mA_{\{j\}}- \mT \mB_{\{j\}}' - \mB_{\{j\}}\mT' + \mT \mC_{\{j\}} \mT' )\right] \\
	&- \frac{n}{2}\log|\mH| -\frac{1}{2} \tr \left[ \mH^{-1} ( \mD_{\{j\}} - \mZ \mE_{\{j\}}' - \mE_{\{j\}} \mZ' + \mZ \mF_{\{j\}} \mZ') \right], \nonumber
\end{align}
where in our case $\mQ=\mI$, while $\mT$, $\mZ$ and $\mH$ are functions of the vector of unknown parameters $\vtheta$ and a possible dependence of the initial conditions for $\valpha_0$ on $\vtheta$ has been discarded for simplicity. The conditional moment matrices $\mA_{\{j\}}$, $\mB_{\{j\}}$, \ldots, are given in appendix \ref{app:a2} and can be computed by a single run of a state smoothing algorithm \citep[][section 4.4]{DurKoo2012} based on the system determined by $\vtheta_{\{j\}}$.

Rather than carrying out the full maximization of $Q(\vtheta, \vtheta_{\{j\}})$ at each step, we obtain a computationally simpler modified algorithm. To this end, we partition the vector of unknown parameters as $\vtheta'=(\vtheta^{(1)'},\vtheta^{(2)'})$ where $\vtheta^{(1)'} = (\vd',\vlambda',\varphi')$, $\vlambda$ contains the unknown elements in $\mLambda$, $\varphi$ holds the unobserved parameters for $\vu_t$ in $\mT^{(2,2)}$ and $\mZ^{(2)}$, while the noise variance parameters in $\mH$ are collected in $\vtheta^{(2)}$. First, the expectation / conditional maximization (ECM) algorithm described by \citet{MenRub1993} in our setup amounts to a conditional optimization over $\vtheta^{(1)}$ for given variance parameters $\vtheta^{(2)}_{\{j\}}$ and optimization over $\vtheta^{(2)}$ for given $\vtheta^{(1)}_{\{j\}}$. Second, as suggested by \citet{WatEng1983}, the optimization over $\vtheta^{(1)}$ is not finalized for each $j$, but rather a single Newton step is implemented for each iteration of the procedure. Neither of these departures from the basic EM algorithm hinders reasonable convergence properties.
%

A Newton step in the estimation of $\vtheta^{(1)}$ for given $\vtheta^{(2)}_{\{j\}}$ yields the estimate in the $(j+1)$-th step
\beq \label{eq:em_tz}
\vtheta^{(1)'}_{\{j+1\}} = (\mXi_{\{j\}}'  \mG_{\{j\}} \mXi_{\{j\}})^{-1} \mXi_{\{j\}}'  (\vg_{\{j\}}- \mG_{\{j\}} \vxi_{\{j\}}).
\eeq
The derivation of \eqref{eq:em_tz} and expressions for $\mXi_{\{j\}}$, $\vxi_{\{j\}}$, $\vg_{\{j\}}$ and $\mG_{\{j\}}$ can be found in appendix \ref{app:a2}. Finally, the free variance parameters of $\mH$, collected in $\vtheta^{(2)}$, are estimated using the derivative of  $Q(\vtheta,\vtheta_{\{j\}})$ with respect to $\mH$; see \eqref{eq:deriv_q_2}. The estimate is given by the corresponding elements of
\[
\frac{1}{n}\mL_{\{j\}} := \frac{1}{n} \E_{\vtheta_{\{j\}}}\sum_{t=1}^n \vepsi_t \vepsi_t' = \frac{1}{n}(\mD_{\{j\}} - \mZ \mE_{\{j\}}' - \mE_{\{j\}} \mZ' + \mZ \mF_{\{j\}} \mZ').
\]

For using gradient-based methods in later steps of the maximization, the likelihood score can be obtained with only one run of a state smoothing algorithm. This has been shown by \citet{KooShe1992}, who draw on the result
\[
\left.\frac{\partial Q(\vtheta,\vtheta_{\{j\}})}{\partial \vtheta}\right|_{\vtheta_{\{j\}}} = \left.\frac{\partial l(\vtheta)}{\partial \vtheta}\right|_{\vtheta_{\{j\}}},
\]
where $l(\vtheta)$ denotes the Gaussian log-likelihood of the model. Evaluation of the score for our model can therefore be based on \eqref{eq:deriv_q_1} and \eqref{eq:deriv_q_2}.

An estimate of the covariance matrix can be computed 
using an analytical expression for the information matrix. Denoting by $\vv_t$ and $\mF_t$ the model residuals and forecast error variances obtained from the Kalman filter, the $i$-th element of the gradient vector for observation $t$ is given by
\beq
\frac{\partial l_t(\vtheta)}{\partial \theta_i} = -\frac{1}{2} \tr \left[ \Big( \mF_t^{-1}\frac{\partial \mF_t}{\partial \theta_i} \Big)(\mI - \mF_t^{-1}\vv_t \vv_t') \right] + \frac{\partial \vv_t'}{\partial \theta_i} \mF_t^{-1} \vv_t,
\eeq \label{eq:gradient}
while the $ij$-th element of the information matrix ${\cal \mI}(\vtheta)$ is
\beq \label{eq:fisher}
{\cal I}_{ij}(\vtheta) = \frac{1}{2} \sum_{t=1}^n \tr \left[\mF_t^{-1} \frac{\partial \mF_t}{\partial \theta_i}\mF_t^{-1}\frac{\partial \mF_t}{\partial \theta_j}\right] + \E_{\vtheta} \left[ \sum_{t=1}^n  \frac{\partial \vv_t'}{\partial \theta_i}\mF_t^{-1}\frac{\partial \vv_t}{\partial \theta_j} \right];
\eeq
see \citet[][section 3.4.5]{Har1991}. To obtain a feasible estimator $\hat{{\cal \mI}}(\hat{\vtheta})$, either the expectation term in \eqref{eq:fisher} is omitted, as suggested by \citet{Har1991}, or the techniques of \citet{CavShu1996} may be used to compute the exact Fisher information. An estimate of the covariance matrix of the estimator is then given by
\beq \label{eq:var_info}
\widehat{\Var}_{\text{info}}(\hat{\vtheta}) = \hat{{\cal \mI}}(\hat{\vtheta})^{-1},
\eeq
or by the sandwich form
\beq \label{eq:var_sand}
\widehat{\Var}_{\text{sand}}(\hat{\vtheta}) = \hat{{\cal \mI}}(\hat{\vtheta})^{-1}\left[ \sum_{t=1}^n \left.\frac{\partial l_t(\vtheta)}{\partial \vtheta} \right|_{\hat{\vtheta}} \left.\frac{\partial l_t(\vtheta)}{\partial \vtheta'} \right|_{\hat{\vtheta}} \right] \hat{{\cal \mI}}(\hat{\vtheta})^{-1},
\eeq
which is robust to certain violations of the model assumptions; see \citet{Whi1982}.

The asymptotic theory for maximum likelihood estimation in the fractionally cointegrated state space setup with integration orders $d \in [0, 1.5)$ is derived in \cite{HarTscWeb2019} for an exact representation of \eqref{eq:FC_frac}. As shown there, the approximation error of the Kalman filter that results from ARMA approximations can be calculated via \eqref{approximation_error} and is $\E_\vtheta(\tilde{x}_t - x_t | y_1,...,y_{t-1})$. Hence, it is measurable given the $\sigma$-field generated by $y_1,...,y_{t-1}$, such that an approximation-corrected estimator can be constructed \citep[][Corollary 2.3]{HarTscWeb2019}. For this estimator, consistency and asymptotic (mixed) normality is shown. While the approximation-corrected maximum likelihood estimator is computationally feasible for long time series (i.e.\ $n$ large), it is limited to low-dimensional $\vy_t$. Thus, especially for models where $\vy_t$ typically holds a large number of observable variables, e.g.\ factor models, the proposed ARMA approximations provide a computationally feasible parametrization of state space models. We compare the performance of the maximum likelihood estimator for our approximate state space model with the approximation-corrected maximum likelihood estimator of \cite{HarTscWeb2019} in a Monte Carlo study in section \ref{ssec:mc1}, where it will become clear that the mean squared error of the approximate estimator for $d$ converges to the mean squared error of the exact estimator as $n$ increases.

Our estimation approach can be straightforwardly generalized to additional situations of great practical relevance. To include a treatment of further components causing nonstationarity such as deterministic trends or exogenous regressors, one can use diffuse initialization of one or more of the states which may be based on \citet{Koo1997}. While we have discussed maximum likelihood estimation under a setting where all data in $\vy_t$ are available, our algorithms can be generalized for arbitrary patterns of missing data using the approach of \citet{BanMod2012}. For very high-dimensional datasets, the computational refinements of \citet{JunKoo2014} may be used. 
For common trends of similar persistence, nonparametric averaging methods may turn out to be useful \citep[cf.\ e.g.][]{ErgRod2016}.

%% file: Sec4_MC.tex
\section{A Monte Carlo study} \label{sec:mc}

We study the performance of the described methods for a number of stylized processes which are nested in the general setup \eqref{eq:FC}. The simulation study is designed to answer several questions. Firstly, we assess whether the finite-order ARMA approximation of the state space system performs well as compared to other parametric or semiparametric approaches. Secondly, we assess the feasibility of joint estimation of memory parameters and cointegration vectors in bivariate fractional systems with and without polynomial cointegration, again considering popular semiparametric approaches as benchmarks. Thirdly, the precision of cointegration estimators is studied in case of several cointegration relations of different strengths and for higher dimensions of the observed time series.

For each specification, we simulate $R=1000$ replications and estimate the models using semiparametric estimates for $\vd$ from the exact local Whittle estimator as starting values for maximum likelihood estimation. The coefficients of the unobserved components can be recovered via the variance of the fractionally differenced observables, since the disturbance terms are standardized. The precision of the estimators is assessed by the root mean squared error (RMSE) criterion or the bias or median errors of the parameter estimators, of state estimates or of out-of-sample forecasts. We vary over different sample sizes $n \in \{250, 500, 1000\}$ which cover relevant situations in macroeconomics and finance.

\subsection{Finite state approximations in a univariate setup} \label{ssec:mc1}

As the simplest stylized setup of our model, we first assess the fractional integration plus noise case, which has been studied in a stationary setup, e.g., by \citet{GraMag2012}. For mutually independent $\xi_t$ and $\epsi_s$, the data generating process is given by
\begin{align} \label{eq:DGP1}
y_t &= \Lambda x_t + \epsi_t, \qquad t=1,\ldots,n, \\
\Delta^{d}x_t &= \xi_t, \quad \xi_t \sim N\!I\!D(0,1), \quad \epsi_t \sim N\!I\!D(0,1). \nonumber
\end{align}
The fractional integration plus noise model is a special case of \eqref{eq:FC} where $\mLambda = \sqrt{q}$, $\vu_t = \varepsilon_t$, $\mathrm{Var}(\varepsilon_t):=h= 1$, and $\xi_t$, $\varepsilon_t$ are independent. For the signal-to-noise ratio we consider $q \in \{0.5, 1, 2\}$, while the memory parameters $d \in \{ 0.25, 0.5, 0.75\}$ cover cases of asymptotically stationary and nonstationary fractional integration. We estimate the free parameters $d$, $q$ and the noise variance $h$ by maximum likelihood using the state space approach.

We apply different approximations to avoid an otherwise $n$-dimensional state process. Firstly, the ARMA($v$,$w$) approximation given by \eqref{eq:arma_1} and \eqref{eq:arma_2} is considered, setting $v=w \in \{2,3,4\}$. The corresponding estimators are denoted as $\hat{d}_{v,w}$ in the result tables. Secondly, we assess truncations of the autoregressive representation of the fractional process at $m=20$ and $m=50$ lags 
as suggested in \citet[section 4.2]{Pal2007}, 
and label these estimators $\hat{d}_{A\!R20}$ and $\hat{d}_{A\!R50}$, respectively. Thirdly, moving average representations as proposed in 
\cite{ChaPal1998} are used, also with a truncation at $m=20$ and $m=50$ lags ($\hat{d}_{M\!A20}$ and $\hat{d}_{M\!A50}$). Furthermore, we employ the exact local Whittle ($\hat{d}_{E\!W}$) estimator of \citet{ShiPhi2005} as well as the univariate exact local Whittle approach ($\hat{d}_{U\!E\!W}$) as defined by \cite{SunPhi2004}, which accounts for additive $I(0)$ perturbations. For both semiparametric estimators of the fractional integration order, we use $m=\lfloor n^{0.65} \rfloor$ Fourier frequencies as a common pragmatic choice. Using other typical values such as $n^j$, $j\in \{0.45, 0.5, 0.55\}$ would not change the results qualitatively, but $n^{0.65}$ is the best choice in most settings considered here. Finally, to grasp the performance of the exact maximum likelihood estimator and to compare our approximate approach with it, we also include the approximation-corrected maximum likelihood estimator of \cite{HarTscWeb2019}, which corrects for the approximation error induced by ARMA(3,3) approximations. 

The root mean squared errors of estimates of $d$ for this setup are shown in table \ref{tab:dgp1_rmse}. Not surprisingly, for this stylized process with only three free parameters, the parametric approaches clearly outperform the semiparametric Whittle estimators. For the EW approach, the performance gets worse for more volatile noise processes (lower $q$), which is not the case for the UEW estimator. The bias of the EW estimator is negative due to the additive noise; see table \ref{tab:dgp1_bias} and also \citet{SunPhi2004}. In contrast, the UEW estimator is positively biased, independently of $q$. Overall, it has inferior estimation properties, so that we do not show the UEW results for the other data generating processes.

Focusing on the state space approximations, we find that the ARMA approach for $v,w\geq 3$ is always among the best approaches. Overall, the ARMA(3,3) and ARMA(4,4) approximations exert a very similar performance, and their relative performance does not seem to depend on the specification of $d$ and $q$. The truncation methods, in contrast, show mixed results. The moving average approximation tends to dominate the autoregressive one for smaller $d<0.5$, which mirrors the conclusion from \citet{GraMag2012} in their stationary setting. However, we find that the autoregression is better whenever nonstationary $d\geq 0.5$ or higher signal-to-noise ratios are considered. 


As expected, the exact maximum likelihood estimator of \cite{HarTscWeb2019} outperforms the approximation methods for most parameter settings. Considering the computational costs which are about 10 times higher than for the ARMA-approximations with $n = 250$, and about 250 times higher with $n=1000$, the improvements are moderate, however. The median improvement in RMSE over the ARMA(3,3) across parameter setups is 7.7\%. As the most extreme scenario, the RMSE can be reduced from 0.132 in the ARMA(3,3) method to 0.102 by the exact estimator for $n =250$, $q=0.5$, $d = 0.25$. As the signal-to-noise ratio $q$ increases, benefits from the approximation-corrected estimator get smaller, and also an increase in the integration orders lowers the benefits from the approximation-corrected estimator. But most interestingly, the RMSE of the ARMA approximations converges to the RMSE of the approximation-corrected estimator as $n$ increases. This indicates that for long time series, where the approximation-correction is particularly costly, it may not even be required. 


Directing attention to table \ref{tab:dgp1_bias} again, we find that the bias for the ARMA approach for $v,w\geq 3$ does not contribute significantly to the estimation errors. Often, it does not appear until the third decimal place. The bias is generally small also for the truncation approaches, but there exist some situations where it is noticeable, mostly for larger $d$. There, larger sample sizes even tend to increase the bias, while higher truncation lags do not always lessen the problem.

We investigate if the results carry over to estimation precision of the fractional components and to forecasting performance of the different approaches. This seems to be the case as table \ref{tab:dgp1_alpha_pred} shows. We restrict attention to the medium signal-to-noise case $q=1$ and apply only the state space approaches. In the upper panel, the fractional component $x_t$ is estimated by a Kalman smoother and the RMSE averages across all in-sample observations and iterations. We find rather small differences between the approaches, especially for small $d$, while ARMA(4,4) and ARMA(3,3) dominate the approximation-based methods in each constellation. The exact method is only slightly superior.
The same holds for the forecasting performance, where 1- to 20-step ahead forecasts are evaluated against realized trajectories, again by their RMSE averaging both across horizons and iterations. The differences between the approaches are only slightly more pronounced than above, and again, ARMA(3,3) and ARMA(4,4) are very close to the best-performing exact method. Interestingly, with the low signal-to-noise ratio (not shown in the table), each approach does a poorer job to recover the underlying fractional component, and neither is able to appropriately separate the fractional from the noise component in any case. 

In sum, we find good performance of the ARMA approximations. The ARMA(3,3) approach appears sufficient in typical empirical applications. This finding is very appreciable in light of the great reduction in computational effort: A fractional component is represented by 4 states, rather than by 50 in a truncation setup with inferior performance, while an approximation-corrected approach has higher computational costs especially for $n=1000$ even in this very simple setup. Both these alternatives can easily become impractical in more complex situations.

Overall, the differences between the approximations account for a small fraction of the overall estimation uncertainty, even in this stylized setting with high overall estimation precision. Also the benefits of the approximation-corrected approach are limited. Together with the finding of accurate ARMA approximations in section \eqref{sec:arma}, this suggests that the need of approximations might not be a serious obstacle to the state space modelling of fractional unobserved components.

\subsection{A basic fractional cointegration setup} \label{ssec:mc2}

The performance of the state space approach in estimating fractionally cointegrated systems is studied in a bivariate process with short-run dynamics,
\begin{align} \label{eq:DGP2}
y_{1t} &= \Lambda_{11} x_t + \Gamma_{11} z_{1t}, \qquad
y_{2t} = \Lambda_{21}x_t+ \Gamma_{21} z_{1t} + \Gamma_{22} z_{2t},\\
\Delta^{d} x_t &= \xi_t, \quad \xi_t \sim N\!I\!D(0,1), \nonumber \\
(1- \phi_i L) z_{it} & = \zeta_{it}, \quad \zeta_{it} \sim N\!I\!D(0,1), \quad i=1,2, \quad t=1,\ldots,n,  \nonumber
\end{align}
with $\Lambda_{11}=\Lambda_{21}=1$, $\Gamma_{11}= \Gamma_{22} =c$, $\Gamma_{21} = c \cdot e$, and $\phi_1 = \phi_2 = 0.5$. This implies that the true cointegration vector $\mB := \mLambda_\perp = (1, -1)'$, $\mB' \begin{pmatrix} y_{1t} & y_{2t}\end{pmatrix}' \sim I(0)$, where the first entry was normalized to one.  
Again the innovations are mutually independent. Note that $u_{1t}=c z_{1t}$, $u_{2t}=(c\cdot e)z_{1z}+cz_{2t}$, which allows for an interpretation of \eqref{eq:DGP2} as a fractionally cointegrated setup with cross- and autocorrelated short-run dynamics. We vary over values of the fractional integration order $d \in \{ 0.25, 0.5, 0.75\}$. The perturbation parameter $c\in\{0.5,1,2\}$ controls the signal-to-noise ratio and short-memory correlation between the processes is introduced, which will be governed by different values of $e \in \{0, 0.5, 1\}$. Cases where $\Gamma_{11} \neq \Gamma_{21}$ could be considered straightforwardly. 

Here and henceforth, we apply the ARMA(3,3) approximation for maximum likelihood estimation of the unknown model parameters. In the current setup, the latter consist of the eight entries in $\vtheta' = (d, \phi_1, \phi_2, \mLambda_{11}, \mLambda_{21}, \mGamma_{11},\mGamma_{21},\mGamma_{22})$, where $\mGamma_{ij}$ is the loading of $z_{jt}$ on $y_{it}$, while the variance parameters are normalized to achieve identification. Starting values for the AR parameters are obtained by fitting an autoregressive model for the difference $y_{1t}-y_{2t}$. To contrast the properties to standard semiparametric approaches again, we apply the EW estimator componentwise to the univariate processes and investigate the mean of the univariate estimates. For the cointegration relation we apply the narrow-band least squares estimator which has been studied by \citet{RobMar2001} in the nonstationary single equation case and by \citet{Hua2009} in a setup with cointegration subspaces \citep[for details on cointegration subspaces, see][]{HuaRob2010, HarWei2018a}. We follow the literature which suggests to use a small number of frequencies and choose $\lfloor n^{0.3} \rfloor$, amounting to 5, 6 and 7 frequencies for our sample sizes.

Since the cointegration vectors are not identified without further restrictions, we investigate the angle $\vartheta$ between true and estimated cointegration spaces. \citet{Nie2010}
provides an expression for the sine of this angle, which is
given 
in our framework by
\beq \label{eq:angle}
\sin(\vartheta) = \frac{\tr(\mLambda\hat{\mB})}{\|\mLambda\| \|\hat{\mB}\|},
\eeq
where $\hat{\mB}$ is an estimated cointegration matrix and $\|  \mA  \|$ is the Euclidean norm of $\mA$. In the current bivariate setup with one cointegration relation, we have $\hat{\mB} = \hat{\mLambda}_{\bot}$ for the maximum likelihood estimator and $\hat{\mB}_{N\!B} = (1, -\hat{\beta}_{N\!B})'$ for the narrow-band least squares estimator $\hat{\beta}_{N\!B}$ applied to $y_{1t} = \beta y_{2t} + \text{error}$. Values of $\sin(\vartheta)$ closer to zero indicate preciser estimates and thus we compute the corresponding root mean squared error criterion as the square root of $\frac{1}{R}\sum_{i=1}^R \sin(\vartheta^i)^2$ in what follows. To get some intuition for the bivariate case, estimating a true value $\mB= (1,-1)'$ by $\hat{\mB} = (1,-1.1)'$ would result in a loss of $\sin(\vartheta) \approx 0.05$.

In table \ref{tab:dgp2_rmse} we show root mean squared errors for memory parameters ($\hat{d}^{M\!L}$ and $\hat{d}^{E\!W}$) and evaluate estimated cointegration spaces (by $\vartheta^{M\!L}$ and $\vartheta^{N\!B}$) applying either the maximum likelihood or the semiparametric technique, respectively. Consider the case $e = 0$ first. Regarding the memory estimators, we find relatively large errors for this data generating process, with root mean squared errors frequently around 0.2 or larger, most prominently when the variances of the short-memory processes are large ($c=2$). The Whittle estimator often performs better than maximum likelihood, especially for smaller $c$ and $d$ and in smaller samples.

For estimating the cointegration space, however, the state space approach appears worthwhile and outperforms narrow band least squares in most constellations. Not surprisingly, strong cointegration relations ($d=0.75$) are precisely estimated, as is cointegration with small short-memory disturbances ($c=0.5$). While the relative merits of maximum likelihood are unchanged for different cointegration strengths, we find that strong perturbations are better captured by the state space estimators. For $c=2$, the RMSE of the semiparametric approach exceeds the parametric RMSE by about $70\%$ in some cases.

Short memory correlation as introduced through $e>0$ overall decreases the precision of the memory estimators. Interestingly, however, the performance of the cointegration estimators improves when $e>0$ is considered. This is the case for both the maximum likelihood and the narrow band approach. To gain some insights into this finding, we assess the typical signed errors of the cointegration estimates. To this end, we consider a normalization of the cointegration vectors as $(1, -\beta)$, and assess estimated $\beta$ for both approaches. 
Note that the narrow-band least squares estimator estimates $\beta_{NB}$ directly, whereas $\beta_{ML}$ is computed via $\beta_ {ML}=-\hat{\Lambda}_{21}/\hat{\Lambda}_{11}$. For $\hat{\Lambda}_{11}$ small, the estimator becomes imprecise. Therefore, it is informative to compute an outlier-robust measure of the typical signed deviation. The median errors ($\median_i(\hat{\beta}_j^i)-\beta_j$) for this data generating process are shown in table \ref{tab:dgp2_mede}. 

The typical deviations for the narrow band estimates exert a negative median bias of the estimates. A positive correlation between the short-memory components appears to work in the opposite direction so that the negative bias is reduced. In contrast, we find that the maximum likelihood estimators are essentially median-unbiased. Here, correlation between the short-memory components may improve the distinction between short and long-memory components and hence reduce variability.

\subsection{Correlated fractional shocks and polynomial cointegration} \label{ssec:mc3}


A further simulation setup extends the model setup \eqref{eq:FC} by introducing contemporaneously correlated $\vxi_t \sim \mathrm{NID}(0, \mS)$, and also allowing for polynomial cointegration through perfectly correlated $\vxi_t$. Polynomial cointegration refers to a situation where lagged observations nontrivially enter a cointegration relation; see \citet{GraLee1989} as well as \citet[][section 4]{Joh2008} for nonfractional and fractional treatments, respectively. 
To motivate polynomial cointegration in terms of our model, assume for simplicity $d_1 > d_2 > ... > d_s$. Let $\Lambda^{(1:(m-1))}$ hold the first $m-1$ columns of $\mLambda$, and let $\mLambda_\perp^{(1:(m-1))}$ be its orthogonal complement. Then $\mLambda_\perp^{(1:(m-1))'}\vy_t \sim I(d_{m})$ annihilates the first $m-1$ common unobserved components $x_{1t}, ..., x_{m-1,t}$. If a vector $\vgamma$ exists, such that $\vgamma' (\vy_t' \mLambda_\perp^{(1:(m-1))} ,\ \Delta^b y_{it})'$ is integrated of a lower order than $d_{m}$ for any $b$ and any $y_{it} \sim I(d_k)$, $k \in \{1,...,m-1\}$, then polynomial cointegration occurs. Whenever $| \mathrm{Cor}(\xi_{kt}, \xi_{mt})| = 1$,  also $x_{mt}$ and $\Delta^{d_k-d_m}x_{kt}$ are perfectly correlated, and hence there exists a linear combination $\vgamma' (
\vy_t' \mLambda_\perp^{(1:m)},\ \Delta^{d_k - d_m} y_{it}
)'$ with a smaller integration order than $d_{m}$. 

We consider
\begin{align} \label{eq:DGP3}
y_{1t} &= \Lambda_{11}x_{1t}  +  \Lambda_{12} x_{2t} + \epsi_{1t}, \qquad
y_{2t} = \Lambda_{21}x_{1t} + \Lambda_{22} x_{2t} + \epsi_{2t},\\
\Delta^{d_i}x_{it}&= \xi_{it}, \quad
\xi_{it} \sim N\!I\!D (0,1), \quad \Corr(\xi_{1t},\xi_{2t})=r,  \nonumber \\
\epsi_{it} &\sim N\!I\!D (0,h_{ii}), \quad h_{ii}=1, \quad  i=1,2, \quad t=1,\ldots,n,  \nonumber
\end{align}
where $\Lambda_{11}=\Lambda_{21}=1$, $\Lambda_{12} = - \Lambda_{22} = a$, $d_1 > d_2$, and where we drop the assumption of orthogonal long-run shocks and allow for $\mathrm{Var}(\xi_t)=\mQ \neq \mI$.
Correlation between the innovations to the fractional processes is introduced through the parameter $r$. Besides the standard setting $r=0$, we refrain from the assumption of independent components for $r=0.5$, while $r=1$ amounts to $\xi_{1t}=\xi_{2t}$ which is the case of polynomial cointegration since there is a second nontrivial cointegration relation in $(y_{1t}, y_{2t}, \Delta^{d_1-d_2} y_{2t})'$. Combinations of $d_2\in \{ 0.2, 0.4\}$ and $d_1\in \{ 0.6, 0.8\}$ contrast relatively weak and strong cases of cointegration, while the importance of the component $x_{2t}$ varies with $a\in \{0.5,1,2\}$. We treat $\vtheta = (d_1, d_2, \mLambda_{11},\mLambda_{21},\mLambda_{12}, \mLambda_{22},r, h_{11}, h_{22})'$ as free parameters, but also investigate estimates imposing the singularity $r=1$ when it is appropriate. Starting values for the fractional integration orders are obtained via the exact local Whittle estimator as in the preceding sections, where we consider the sum and the difference of $y_{1t}$ and $y_{2t}$ to estimate $d_1$ and $d_2$. Initial values for $r$ are obtained from the covariance of the fractionally differenced processes $\mathrm{Cov}(\Delta^{d_2}(y_{1t}-y_{2t}), \Delta^{d_1}(y_{1t}+y_{2t}))$. 

Consider the results for $r=0.5$ first. The root mean squared errors, shown in table \ref{tab:dgp3_r05}, include estimators of cointegration spaces as above (evaluated by $\vartheta_1^{M\!L}$ and $\vartheta_1^{N\!B}$ in the table). Now, there are two memory parameters to be estimated either by maximum likelihood ($\hat{d}_1^{M\!L}$ and $\hat{d}_2^{M\!L}$) or by the Whittle approach ($\hat{d}_1^{E\!W}$ and $\hat{d}_2^{E\!W}$). Semiparametric estimates of $d_2$ are obtained from the narrow band least squares residuals. The table also contains the maximum likelihood estimate of the correlation parameter $r$ ($\hat{r}^{M\!L}$).

For most parameter settings, we observe that the parametric memory estimators perform satisfactorily. They outperform the semiparametric approach whenever there is strong influence of the $x_{2t}$ components ($a=2$), most pronouncedly in larger samples. 
Also regarding cointegration estimators, higher values of $a$ favor the parametric method. The correlation parameter is estimated with increasing precision in larger samples, while also the strength of the cointegration relation is relevant for this estimator. For $d_1=d_2$, the correlation parameter (and also certain elements of $\mLambda$) would not be identifiable, and hence setups with small difference $d_1-d_2$ are problematic.

For $r=1$, we additionally consider the properties of estimators for the polynomial cointegration relation. To evaluate estimators of the polynomial cointegration spaces, note that the cointegration space leading to the highest memory reduction in $(y_{1t}, y_{2t}, \Delta^{d_1-d_2} y_{2t})'$ is the orthogonal complement of the span of
\beq \label{eq:poly_ci_space}
\bmat \mLambda^{(1)} & \mLambda^{(2)} \\ 0 & \mLambda_{21} \emat,
\eeq
where $\mLambda^{(j)}$ refers to the $j$-th column of $\mLambda$.
This cointegration subspace is estimated replacing all entries in \eqref{eq:poly_ci_space} by their maximum likelihood estimates, where $r=1$ is imposed. For the narrow band least squares estimator, this space is determined by the span of $(1, -\hat{\beta}_1,-\hat{\beta}_2)'$, where the coefficients are narrow band least squares estimates from $y_{1t} = \beta_1 y_{2t} + \beta_2 \Delta^{d_1-d_2} y_{2t} + \text{error}$ with $d_1$ and $d_2$ replaced by local Whittle estimates. Estimators for this second (polynomial) cointegration relation are evaluated analoguously to \eqref{eq:angle} where now \eqref{eq:poly_ci_space} takes the role of $\mLambda$ and the resulting angle is denoted by $\vtheta_2$.

In table \ref{tab:dgp3_r1}, the corresponding root mean squared errors are given. The elementary cointegration space is estimated by the unrestricted estimator (see $\vartheta_1^{M\!L}$) and the restricted estimator (see $\vartheta_1^{R\!M\!L}$, imposing $r=1$) with a very similar precision. This is in accordance with the notably precise estimation of $r$ in this case. The parametric estimators of both cointegration spaces are again better than semiparametric approaches (1) in large samples and (2) when a strong second fractional component is present. Overall, the results suggest that polynomial fractional cointegration analysis is feasible in our setup, while the maximum likelihood approach has reasonable estimation properties at least for larger sample sizes.

\subsection{Cointegration subspaces in higher dimensions} \label{ssec:mc4}

Until now, we have considered one- or two-dimensional processes in our simulations which limits the empirical relevance of the findings so far. We claim that modelling high-dimensional time series constitutes a strength of our approach, at least if suitably sparse parametrizations with factor structures are empirically reasonable. As a second generalisation compared to the previous setups, we consider situations where two or more cointegration relations exist and where these may be of different strength, i.e., where the reduction in memory through cointegration differs among relations. The latter situation has been studied under the label of cointegration subspaces, among others by \citet{HuaRob2010} and \citet{HarWei2018a}.

To assess the performance in this situation, consider the process
\begin{align} \label{eq:DGP4}
y_{it} &= \Lambda_{i1} x_{1t} + \Lambda_{i2} x_{2t} + \epsi_{it},
\\
\Delta^{d_j}x_{jt} &= \xi_{jt}, \quad \xi_{jt} \sim N\!I\!D (0,1),\nonumber \\
\epsi_{it} &\sim N\!I\!D (0,h_{ii}), \quad h_{ii} = 1, \quad j=1,2, \quad i=1,\ldots,p, \quad t=1,\ldots,n, \nonumber
\end{align}
where $\Lambda_{i1}=a$, $\Lambda_{i2} = a \cdot (-1)^{i+1}$ $\forall i = 1,...,p$, and with mutually independent noise sequences. We now vary over the dimension $p \in \{3,10,50\}$, while again combinations of $d_1\in \{ 0.2, 0.4\}$ and $d_2\in \{ 0.6, 0.8\}$ are considered. The parameter $a\in \{0.5,1,2\}$ gives the relative importance of the fractional components and hence plays the role of a signal-to-noise ratio. We estimate $d_j$, $\mLambda_{ij}$, $h_i$ for $j=1,2$ and $i=1,\ldots,p$ as free parameters. Starting values for $d_1$ and $d_2$ are obtained as in section \ref{ssec:mc3}.

Along with the memory estimates, we show results for estimating the $p-1$ cointegration relations reducing the memory from $d_1$ to $d_2$ (the first cointegration subspace) which is evaluated by the angle $\vartheta_1$ between $\mLambda^{(1)}$ and the cointegration matrix estimate $\widehat{\mB}_1$. Additionally, the $p-2$ cointegration relations reducing the memory from $d_1$ to $0$ (the second cointegration subspace) are evaluated by the angle $\vartheta_2$ between $\mLambda$ and $\widehat{\mB}_2$. The cointegration matrices are straightforwardly obtained for the maximum likelihood approach by the orthogonal complements of $\widehat{\mLambda}^{(1)}$ and $\widehat{\mLambda}$, respectively. The narrow-band least squares method estimates cointegration matrices under specific normalizations as above. Estimating the first subspace, we construct $\hat{\mB}_1$ to have free entries $-\hat{\beta}_{2}$, \ldots, $-\hat{\beta}_{p}$ in the first row and a $p-1$ identity matrix below, such that $\beta_{j}$ is obtained from $y_{jt} = \beta_j y_{1t} + \text{error}$ for $j=2,\ldots,p$. In the estimation of the second subspace, we have two free rows in $\hat{\mB}_2$ which are given by $(-\hat{\beta}_{13}$, \ldots, $-\hat{\beta}_{1p})$, and $(-\hat{\beta}_{23}$, \ldots, $-\hat{\beta}_{2p})$, respectively, and can be estimated from $y_{jt} = \beta_{1j} y_{1t} + \beta_{2j} y_{2t} +\text{error}$ for $j=3,\ldots,p$.

In table \ref{tab:dgp4}, results are shown for $a=0.5$ while the other specifications yield qualitatively similar outcomes. The process allows for a precise estimation of both $d_1$ and $d_2$ by maximum likelihood. An increasing dimension $p$ leads to a better estimation by maximum likelihood which is not the case for the Whittle technique. The semiparametric Whittle estimates are obtained by averaging univariate estimates for $d_1$ and using narrow band least squares residuals to estimate $d_2$. Notably, the estimates of $d_2$ hardly improve with larger $n$, which can be explained by a specific shortcoming of the single equation approach: The univariate regression errors may each have integration orders of $d_2$ or lower. In our case, lower orders prevail for $y_{jt} = \beta_j y_{1t} + \text{error}$ with $j$ odd, due to the special structure of $\mLambda$. Knowledge about this specific structure is not exploited by both methods, however, to keep the simulation scenario realistic.

Also regarding the estimation of the cointegration spaces, maximum likelihood is superior. Both parametric and semiparametric estimators have smaller errors for higher dimension, whereas this ``blessing of dimensionality'' is more pronounced for the state space approach. Generally, the ratio between the maximum likelihood RMSE and the semiparametric RMSE decreases for larger $p$.

Not surprisingly, the case with strongest basic cointegration (large difference $d_1 - d_2$, which implies a great reduction of persistence when $x_1$ is projected out) is the one with highest precision in estimating the first cointegration subspace. For estimating the second subspace, a slightly different logic applies, with a larger $d_2$ supporting the estimation. E.g., in the case $d_1 =0.6$ and $d_2=0.4$ a higher precision is achieved than for $d_1 =0.6$ and $d_2=0.2$. Overall, we find that our approach profits from imposing the factor structure which is not the case for the benchmark methods applied in this comparison.

%% file: Sec5_Concl.tex
\section{Conclusion} \label{sec:conclusion}

We have proposed estimation methods for nonstationary unobserved components models which are computationally efficient and provide a good approximation performance. These may be relevant for a wide variety of applications in macroeconomics and finance, as \citet{HarWei2018a} have illustrated. Further work is needed to assess the performance of the methods in different, possibly very high-dimensional, settings.

\section*{Acknowledgements}

The research of this paper has partly been conducted while Roland Weigand was at the University of Regensburg and at the Institute for Employment Research (IAB) in Nuremberg. Very valuable comments by Rolf Tschernig, Enzo Weber, and two anonymous referees, are gratefully acknowledged. Tobias Hartl gratefully acknowledges support through the projects TS283/1-1 and WE4847/4-1 financed by the German Research Foundation (DFG).

%% file: SecA1_App.tex
\appendix

\section{Computational details of the approximating ARMA coefficients} \label{app:a1}
As shown in \eqref{eq:arma_2} the ARMA approximation of a fractionally integrated process for a given integration order $d$ is defined as the set of ARMA parameters that minimize \eqref{eq:MSE_d}. The minimization problem has a unique solution for $p+q \leq n$.
We conduct the optimization \eqref{eq:arma_2} to obtain ARMA approximations of a fractional process over an appropriate, possibly nonstationary, range of $d$. For $d<1$, we impose stability of the autoregressive polynomial, while imposing unit roots is found to enhance the numeric stability of the optimization for $d\geq 1$. In order to achieve numerically well-behaved optimizations, we work with transformed parameters and then re-transform them when the optimum is reached. First, the stable autoregressive and moving average parts are individually mapped to the space of partial autocorrelations so that they take values in $(-1, 1)$; see \citet{BarSch1973} and \citet{Vee2012}. Then, we apply Fishers z-transform $z=0.5[\log(1+x)-\log(1-x)]$ to obtain an unconstrained optimization problem. For a given sample size $n$, we carry out an optimization for each value on a grid for $d$. We smooth the values using cubic regression splines before the result is re-transformed to the space of ARMA coefficients. In this way, we obtain a continuous and differentiable function $\hat{\varphi}_n(d)$. Whenever discontinuities occur in the space of transformed parameters (as for $d=1$), we enforce a smooth transition between segments of $\hat{\varphi}_n(d)$ by the sine function. All computations in this paper are conducted using R \citep{R2018}.

\section{Details on the EM Algorithm} \label{app:a2}

In this appendix, all necessary expressions for the computation of the EM algorithm will be given. The log-likelihood where the unobserved state process $\valpha_t$ is assumed known is called the complete data log likelihood and given by
\begin{align*}
	l(\vtheta;\{\vy_t,\valpha_t\}_{t=1}^n) =& -\frac{n}{2} \log|\mQ|  - \frac{1}{2} \tr\Big[\mR\mQ^{-1}\mR' \sum_{t=2}^n(\valpha_{t}-\mT\valpha_{t-1})(\valpha_{t}-\mT\valpha_{t-1})'\Big] \\
	&- \frac{n}{2}\log|\mH| -\frac{1}{2} \tr\Big[\mH^{-1} \sum_{t=1}^n(\vy_t-\mZ\valpha_{t})(\vy_t-\mZ\valpha_{t})'\Big].
\end{align*}
The expectation of the complete data likelihood, with expectation evaluated at parameters $\vtheta_{\{j\}}$, is denoted by $Q(\vtheta,\vtheta_{\{j\}})$ and given by \eqref{eq:em_q}. The terms involving expectations of the (partially unobserved) data and its cross-moments are
\begin{align*}
	\mA_{\{j\}}&:=  \E_{\vtheta_{\{j\}}}\Big[ \sum_{t=2}^n \valpha_t \valpha_t' \Big] =  \sum_{t=2}^n \hat{\valpha}_t \hat{\valpha}_t' + \sum_{t=2}^n \mV_{t,t}, \\
	\mB_{\{j\}} &:=  \E_{\vtheta_{\{j\}}}\Big[ \sum_{t=2}^n \valpha_t \valpha_{t-1}' \Big] =  \sum_{t=2}^n \hat{\valpha}_t \hat{\valpha}_{t-1}' + \sum_{t=2}^n \mV_{t,t-1}, \\
	\mC_{\{j\}} &:=  \E_{\vtheta_{\{j\}}}\Big[ \sum_{t=2}^n \valpha_{t-1} \valpha_{t-1}' \Big] =  \sum_{t=2}^n \hat{\valpha}_{t-1} \hat{\valpha}_{t-1}' + \sum_{t=2}^n \mV_{t-1,t-1}, \\
	\mD_{\{j\}} &:=  \E_{\vtheta_{\{j\}}}\Big[ \sum_{t=1}^n \vy_{t} \vy_{t}' \Big] =  \sum_{t=1}^n \vy_{t} \vy_{t}', \quad
	\mE_{\{j\}} :=  \E_{\vtheta_{\{j\}}}\Big[ \sum_{t=1}^n \vy_{t} \valpha_t'\Big] =  \sum_{t=1}^n \vy_{t} \hat{\valpha}_t', \\
	\mF_{\{j\}} &:=  \E_{\vtheta_{\{j\}}}\Big[ \sum_{t=1}^n \valpha_t \valpha_t' \Big] =  \sum_{t=1}^n \hat{\valpha}_t \hat{\valpha}_t' + \sum_{t=1}^n \mV_{t,t}.
\end{align*}
Here, $\hat{\valpha}_t=\E_{\vtheta_{\{j\}}}[\valpha_t]$ and $\mV_{t,s}=\E_{\vtheta_{\{j\}}}[(\valpha_t-\hat{\valpha}_t)(\valpha_s-\hat{\valpha}_s)']$ can be computed by state smoothing algorithms based on the state space representation for given $\vtheta_{\{j\}}$ \citep[][section 4.4]{DurKoo2012}.

We turn to the derivation of \eqref{eq:em_tz}. For notational convenience we denote the objective function for optimization over $\vtheta^{(1)}$ by $Q_{\{j\}}^{(1)}(\vtheta^{(1)})\equiv Q((\vtheta^{(1)'},\vtheta^{(2)'}_{\{j\}})';\vtheta_{\{j\}})$. To describe the Newton step in the optimization of $Q_{\{j\}}^{(1)}$ in detail, we explicitly state the nonlinear dependence of $\vec(\mT,\mZ)'=(\vec(\mT)',\vec(\mZ)')$ on $\vtheta^{(1)}$ by $\vec(\mT,\mZ) = \vf(\vtheta^{(1)})$ and consider the linearization at $\vtheta_{\{j\}}$,
\beq \label{eq:lin}
\vec(\mT,\mZ) \approx \mXi_{\{j\}}\vtheta^{(1)} + \vxi_{\{j\}},\quad \text{where} \quad \mXi\equiv \frac{\partial \vf(\vtheta^{(1)})}{\partial \vtheta^{(1)'}},
\eeq
$\vxi \equiv \vf(\vtheta^{(1)})-\mXi\vtheta^{(1)}$, and the $\{j\}$ subscript indicates evaluation of a specific expression at $\vtheta_{\{j\}}$. The optimization over $\vtheta^{(1)}$ jointly involves elements in $\mT$ and $\mZ$, since $\vd$ enters the expression of both system matrices and hence, $\mXi$ is not diagonal.

A single iteration of the Newton optimization algorithm is carried out by expanding the gradient around $\vtheta^{(1)}_{\{j\}}$. The gradient is given by
\beq \label{eq:deriv_q_1}
\frac{ \partial Q_{\{j\}}^{(1)}(\vtheta^{(1)})}{\partial \vtheta^{(1)}} =
\frac{ \partial (\vec(\mT)',\vec(\mZ)')}{\partial \vtheta^{(1)}}
\frac{\partial Q_{\{j\}}^{(1)}}{ \partial (\vec(\mT)',\vec(\mZ)')'}
= \mXi' \vec\bmat \frac{\partial Q_{\{j\}}^{(1)}}{ \partial \mT}& \frac{\partial Q_{\{j\}}^{(1)}}{\partial \mZ}\emat,
\eeq
where we drop the function argument of $Q_{\{j\}}^{(1)}(\vtheta^{(1)})$ for notational convenience.
For the derivatives with respect to the system matrices we have
\[
\frac{ \partial Q_{\{j\}}^{(1)}}{\partial \mT} = (\mR \mQ^{-1} \mR')(\mB_{\{j\}}-\mT\mC_{\{j\}})
\qquad \text{and} \qquad
\frac{ \partial Q_{\{j\}}^{(1)}}{\partial \mZ} = \mH^{-1} (\mE_{\{j\}}-\mZ\mF_{\{j\}}),
\]
so that
\[\vec\bmat \frac{\partial Q_{\{j\}}^{(1)}}{ \partial \mT}& \frac{\partial Q_{\{j\}}^{(1)}}{\partial \mZ}\emat
= \bmat \vec(\mR \mQ^{-1} \mR'\mB_{\{j\}}) \\ \vec(\mH^{-1}\mE_{\{j\}}) \emat
-
\bmat
\mC_{\{j\}}' \otimes \mR \mQ^{-1} \mR' & \mzeros
\\ \mzeros & \mF_{\{j\}}' \otimes \mH^{-1}
\emat
\bmat
\vec(\mT) \\ \vec(\mZ).
\emat
\]
Hence, for $\mG_{\{j\}}$ and $\vg_{\{j\}}$ given by
\[
\vg_{\{j\}} = \vec(\mR \mQ^{-1} \mR' \mB_{\{j\}},\mH^{-1} \mE_{\{j\}}), \quad \text{and} \quad
\mG_{\{j\}} =\diag(\mC_{\{j\}}' \otimes \mR \mQ^{-1} \mR',\mF_{\{j\}}' \otimes \mH_{\{j\}}^{-1}),
\]
we obtain the linear expansion
\[
\frac{ \partial Q_{\{j\}}^{(1)}(\vtheta^{(1)})}{\partial \vtheta^{(1)}} \approx \mXi_{\{j\}}'\vg_{\{j\}} - \mXi_{\{j\}}'\mG_{\{j\}}(\mXi_{\{j\}}\vtheta^{(1)}+\vxi_{\{j\}}).
\]
Equating to zero and solving for $\vtheta^{(1)}$ yields \eqref{eq:em_tz}. For the estimation of $\mH$, see \eqref{eq:deriv_q_2}, we define
\beq \label{eq:deriv_l}
\mL_{\{j\}} :=  \E_{\vtheta_{\{j\}}}\Big[ \sum_{t=1}^n \vepsi_t \vepsi_t' \Big] =  \mD_{\{j\}} - \mZ \mE_{\{j\}}' - \mE_{\{j\}} \mZ' + \mZ \mF_{\{j\}} \mZ'.
\eeq
and use
\beq \label{eq:deriv_q_2}
\frac{\partial Q(\vtheta,\vtheta_{\{j\}})}{\partial \mH} =
(\mH^{-1}\mL_{\{j\}} - n\mI)\mH^{-1} - 0.5 \diag((\mH^{-1}\mL_{\{j\}} - n\mI)\mH^{-1})
\eeq
to derive the estimator of the variance parameters.

%% file: SecA2_Tables_new.tex
%
\clearpage
\begin{table}[ht]
	\centering
	\begin{tabular}{rrr|rrrrrrrrrr}
		q & d & n & $\hat{d}_{2,2}$ & $\hat{d}_{3,3}$ & $\hat{d}_{4,4}$ & $\hat{d}_{A\!R20}$ & $\hat{d}_{A\!R50}$ & $\hat{d}_{M\!A20}$ & $\hat{d}_{M\!A50}$ & $\hat{d}_{E\!W}$&$\hat{d}_{U\!E\!W}$ & $\hat{d}_{exact} $\\
		\hline
.5 & .25 & 250 & .130 & .132 & .132 & .131 & .130 & .123 & .122 & .161 & .414 & .102 \\ 
&      & 500 & .077 & .075 & .075 & .075 & .075 & .075 & .074 & .138 & .324 & .069 \\ 
&      & 1000 & .052 & .050 & .050 & .051 & .050 & .051 & .050 & .119 & .260 & .048 \\ 
& .50 & 250 & .110 & .109 & .106 & .122 & .114 & .106 & .109 & .191 & .305 & .089 \\ 
&      & 500 & .068 & .068 & .068 & .078 & .071 & .071 & .070 & .157 & .212 & .060 \\ 
&      & 1000 & .045 & .045 & .044 & .052 & .047 & .052 & .048 & .125 & .140 & .039 \\ 
& .75 & 250 & .098 & .101 & .100 & .113 & .100 & .150 & .125 & .192 & .223 & .091 \\ 
&      & 500 & .069 & .066 & .066 & .096 & .079 & .108 & .096 & .148 & .157 & .066 \\ 
&      & 1000 & .048 & .044 & .044 & .086 & .058 & .084 & .072 & .108 & .107 & .046 \\ 
1.0 & .25 & 250 & .086 & .086 & .086 & .085 & .085 & .084 & .083 & .132 & .413 & .078 \\ 
&      & 500 & .058 & .057 & .057 & .057 & .057 & .057 & .056 & .111 & .315 & .053 \\ 
&      & 1000 & .040 & .038 & .038 & .039 & .039 & .039 & .038 & .090 & .230 & .037 \\ 
& .50 & 250 & .077 & .078 & .078 & .086 & .081 & .082 & .078 & .145 & .279 & .072 \\ 
&      & 500 & .056 & .054 & .054 & .058 & .057 & .059 & .057 & .118 & .188 & .049 \\ 
&      & 1000 & .038 & .036 & .036 & .042 & .038 & .042 & .040 & .089 & .125 & .033 \\ 
& .75 & 250 & .076 & .075 & .075 & .081 & .075 & .114 & .096 & .143 & .200 & .074 \\ 
&      & 500 & .057 & .054 & .054 & .066 & .055 & .086 & .084 & .111 & .142 & .054 \\ 
&      & 1000 & .044 & .037 & .036 & .068 & .044 & .059 & .069 & .079 & .098 & .038 \\ 
2.0 & .25 & 250 & .072 & .072 & .072 & .071 & .072 & .068 & .068 & .114 & .402 & .065 \\ 
&      & 500 & .049 & .048 & .048 & .048 & .048 & .047 & .047 & .094 & .308 & .044 \\ 
&      & 1000 & .033 & .032 & .032 & .033 & .032 & .033 & .032 & .072 & .197 & .031 \\ 
& .50 & 250 & .067 & .066 & .066 & .075 & .069 & .071 & .067 & .118 & .257 & .062 \\ 
&      & 500 & .049 & .046 & .046 & .051 & .049 & .052 & .050 & .096 & .178 & .043 \\ 
&      & 1000 & .034 & .031 & .031 & .037 & .034 & .037 & .035 & .071 & .117 & .029 \\ 
& .75 & 250 & .067 & .064 & .064 & .069 & .065 & .114 & .088 & .116 & .187 & .064 \\ 
&      & 500 & .052 & .046 & .046 & .057 & .047 & .108 & .080 & .093 & .133 & .046 \\ 
&      & 1000 & .055 & .032 & .032 & .060 & .038 & .098 & .067 & .066 & .093 & .033 \\ 
	\end{tabular} \caption{Root mean squared error (RMSE) for memory parameters in DGP1 \eqref{eq:DGP1}. The columns show maximum likelihood estimators under ARMA($v$,$w$) approximations of the fractional process with $v=w\in\{2,3,4\}$ ($\hat{d}_{v,w}$). Additionally, the truncated AR($m$) representation ($\hat{d}_{A\!Rm}$), and truncated MA($m$) representations ($\hat{d}_{M\!Am}$) are given. Furthermore, we show the exact local Whittle ($\hat{d}_{E\!W}$) and the univariate exact local Whittle estimator ($\hat{d}_{U\!E\!W}$), each with $\lfloor n^{0.65} \rfloor$ Fourier frequencies. Finally, we include results for the approximation-corrected ML estimator ($\hat{d}_{exact}$).} \label{tab:dgp1_rmse}
\end{table}

\begin{table}[ht]
	\centering
	\vspace{-0.6cm}
	\begin{tabular}{rrr|rrrrrrrrrr}
		q & d & n & $\hat{d}_{2,2}$ & $\hat{d}_{3,3}$ & $\hat{d}_{4,4}$ & $\hat{d}_{A\!R20}$ & $\hat{d}_{A\!R50}$ & $\hat{d}_{M\!A20}$ & $\hat{d}_{M\!A50}$ & $\hat{d}_{E\!W}$&$\hat{d}_{U\!E\!W}$ & $\hat{d}_{exact}$\\
		\hline
.5 & .25 & 250 & -.019 & -.019 & -.019 & -.027 & -.025 & -.032 & -.032 & -.125 & .167 & -.014 \\ 
&      & 500 & -.012 & -.010 & -.010 & -.012 & -.015 & -.015 & -.018 & -.112 & .119 & -.004 \\ 
&      & 1000 & -.007 & -.005 & -.005 & -.002 & -.008 & -.005 & -.010 & -.103 & .094 & .005 \\ 
& .50 & 250 & -.006 & -.003 & -.003 & -.008 & -.006 & -.036 & -.034 & -.162 & .095 & -.010 \\ 
&      & 500 & -.011 & -.006 & -.005 & -.005 & -.010 & -.020 & -.030 & -.134 & .040 & -.008 \\ 
&      & 1000 & -.012 & -.003 & -.003 & .011 & -.005 & .002 & -.017 & -.110 & .016 & -.007 \\ 
& .75 & 250 & -.013 & -.005 & -.005 & .012 & -.003 & -.052 & -.066 & -.163 & .035 & -.010 \\ 
&      & 500 & -.016 & -.007 & -.006 & .024 & .002 & -.036 & -.067 & -.123 & .010 & -.010 \\ 
&      & 1000 & -.011 & -.006 & -.003 & .046 & .013 & -.001 & -.052 & -.090 & .004 & -.010 \\ 
1.0 & .25 & 250 & -.011 & -.009 & -.009 & -.017 & -.015 & -.022 & -.020 & -.085 & .207 & -.009 \\ 
&      & 500 & -.009 & -.007 & -.007 & -.010 & -.012 & -.014 & -.014 & -.075 & .139 & -.004 \\ 
&      & 1000 & -.006 & -.004 & -.003 & -.002 & -.006 & -.005 & -.008 & -.067 & .088 & .003 \\ 
& .50 & 250 & -.006 & -.002 & -.002 & -.005 & -.004 & -.036 & -.028 & -.104 & .103 & -.008 \\ 
&      & 500 & -.009 & -.005 & -.004 & -.005 & -.007 & -.024 & -.027 & -.084 & .052 & -.007 \\ 
&      & 1000 & -.009 & -.003 & -.002 & .008 & -.003 & -.005 & -.016 & -.067 & .026 & -.006 \\ 
& .75 & 250 & -.009 & -.006 & -.005 & .009 & -.003 & -.068 & -.068 & -.101 & .048 & -.008 \\ 
&      & 500 & -.005 & -.008 & -.006 & .018 & -.000 & -.054 & -.066 & -.074 & .023 & -.009 \\ 
&      & 1000 & .013 & -.006 & -.003 & .039 & .010 & -.026 & -.054 & -.053 & .015 & -.008 \\ 
2.0 & .25 & 250 & -.004 & -.002 & -.002 & -.011 & -.007 & -.017 & -.013 & -.053 & .224 & -.006 \\ 
&      & 500 & -.007 & -.005 & -.005 & -.008 & -.009 & -.012 & -.012 & -.047 & .148 & -.004 \\ 
&      & 1000 & -.004 & -.002 & -.002 & -.001 & -.004 & -.005 & -.006 & -.041 & .086 & .002 \\ 
& .50 & 250 & -.001 & .001 & .001 & .003 & .001 & -.033 & -.020 & -.061 & .108 & -.006 \\ 
&      & 500 & -.004 & -.003 & -.003 & -.001 & -.003 & -.026 & -.023 & -.049 & .061 & -.006 \\ 
&      & 1000 & -.003 & -.001 & -.001 & .008 & -.000 & -.010 & -.015 & -.038 & .034 & -.005 \\ 
& .75 & 250 & -.000 & -.004 & -.004 & .009 & -.002 & -.068 & -.067 & -.059 & .059 & -.006 \\ 
&      & 500 & .011 & -.006 & -.005 & .017 & -.000 & -.055 & -.066 & -.043 & .033 & -.008 \\ 
&      & 1000 & .041 & -.003 & -.003 & .035 & .009 & -.030 & -.055 & -.030 & .022 & -.005 \\ 
	\end{tabular} \caption{Bias for memory parameters in DGP1 \eqref{eq:DGP1}. The columns show maximum likelihood estimators under ARMA($v$,$w$) approximations of the fractional process with $v=w\in\{2,3,4\}$ ($\hat{d}_{v,w}$). Additionally, the truncated AR($m$) representation ($\hat{d}_{A\!Rm}$), and truncated MA($m$) representations ($\hat{d}_{M\!Am}$) are given. Furthermore, we show the exact local Whittle ($\hat{d}_{E\!W}$) and the univariate exact local Whittle estimator ($\hat{d}_{U\!E\!W}$), each with $\lfloor n^{0.65} \rfloor$ Fourier frequencies.  Finally, we include results for the approximation-corrected ML estimator ($\hat{d}_{exact}$).} \label{tab:dgp1_bias}
\end{table}

\begin{table}[ht]
	\centering
	\begin{tabular}{rr|rrrrrrrr}
		d & n & $\hat{d}_{2,2}$ & $\hat{d}_{3,3}$ & $\hat{d}_{4,4}$ & $\hat{d}_{A\!R20}$ & $\hat{d}_{A\!R50}$ & $\hat{d}_{M\!A20}$ & $\hat{d}_{M\!A50}$& $\hat{d}_{exact}$  \\ 
		\hline
		&& \multicolumn{8}{c}{RMSE of estimated fractional component}\\
		0.25 & 250 & 0.710 & 0.710 & 0.710 & 0.711 & 0.710 & 0.711 & 0.710  & 0.710 \\ 
		& 500 & 0.708 & 0.708 & 0.708 & 0.708 & 0.708 & 0.708 & 0.708  & 0.707 \\ 
		& 1000 & 0.707 & 0.707 & 0.707 & 0.707 & 0.707 & 0.707 & 0.707 & 0.707 \\ 
		0.50 & 250 & 0.700 & 0.700 & 0.700 & 0.701 & 0.700 & 0.704 & 0.701  & 0.699 \\ 
		& 500 & 0.699 & 0.698 & 0.698 & 0.699 & 0.698 & 0.702 & 0.700  & 0.698 \\ 
		& 1000 & 0.699 & 0.697 & 0.697 & 0.698 & 0.697 & 0.701 & 0.699 & 0.697 \\ 
		0.75 & 250 & 0.687 & 0.686 & 0.686 & 0.686 & 0.686 & 0.708 & 0.696  & 0.686 \\ 
		& 500 & 0.687 & 0.685 & 0.685 & 0.685 & 0.685 & 0.709 & 0.697  & 0.685 \\ 
		& 1000 & 0.687 & 0.684 & 0.684 & 0.685 & 0.684 & 0.713 & 0.699 & 0.684 \\  \hline
		&& \multicolumn{8}{c}{RMSE of out-of-sample forecasts}\\
		0.25 & 250 & 1.465 & 1.465 & 1.465 & 1.465 & 1.465 & 1.467 & 1.465  & 1.464 \\ 
		& 500 & 1.463 & 1.463 & 1.463 & 1.464 & 1.463 & 1.465 & 1.464  & 1.463 \\ 
		& 1000 & 1.438 & 1.438 & 1.437 & 1.439 & 1.437 & 1.440 & 1.438 & 1.438 \\ 
		0.50 & 250 & 1.680 & 1.678 & 1.678 & 1.688 & 1.681 & 1.733 & 1.697  & 1.677 \\ 
		& 500 & 1.688 & 1.685 & 1.684 & 1.699 & 1.691 & 1.753 & 1.712  & 1.684 \\ 
		& 1000 & 1.645 & 1.641 & 1.640 & 1.657 & 1.643 & 1.724 & 1.669 & 1.641 \\ 
		0.75 & 250 & 2.262 & 2.261 & 2.261 & 2.282 & 2.260 & 3.004 & 2.524  & 2.260 \\ 
		& 500 & 2.290 & 2.287 & 2.285 & 2.353 & 2.309 & 3.227 & 2.707  & 2.286 \\ 
		& 1000 & 2.201 & 2.191 & 2.191 & 2.264 & 2.212 & 3.394 & 2.677 & 2.191 \\
	\end{tabular}
	\caption{Upper panel: RMSE of estimating the fractional component $x_t$ of DGP1 by the Kalman smoother. The RMSE averages across all in-sample observations and iterations. Lower panel: RMSE of forecasting a realized trajectory of DGP1 out-of-sample. The RMSE averages across all horizons from 1 to 20 and iterations. } \label{tab:dgp1_alpha_pred}
\end{table}

\begin{table}[ht]
	\hspace{-1cm}
	\begin{tabular}{rrr|rrrr|rrrr|rrrr}
		&& & \multicolumn{4}{c|}{$e=0$} & \multicolumn{4}{c|}{$e=0.5$} & \multicolumn{4}{c}{$e=1$} \\
		c & d & n & $\hat{d}^{M\!L}$ & $\hat{d}^{E\!W}$ & $\vartheta^{M\!L}$ & $\vartheta^{N\!B}$ & $\hat{d}^{M\!L}$ & $\hat{d}^{E\!W}$ & $\vartheta^{M\!L}$ & $\vartheta^{N\!B}$ & $\hat{d}^{M\!L}$ & $\hat{d}^{E\!W}$ & $\vartheta^{M\!L}$ & $\vartheta^{N\!B}$ \\
		\hline
.5 & .25 & 250 & .204 & .101 & .212 & .206 & .255 & .106 & .216 & .161 & .277 & .111 & .219 & .156 \\ 
&      & 500 & .154 & .087 & .143 & .161 & .171 & .091 & .132 & .122 & .207 & .095 & .128 & .120 \\ 
&      & 1000 & .086 & .075 & .097 & .125 & .113 & .079 & .088 & .097 & .125 & .086 & .091 & .098 \\ 
& .50 & 250 & .207 & .122 & .094 & .082 & .222 & .126 & .083 & .066 & .238 & .134 & .084 & .063 \\ 
&      & 500 & .126 & .102 & .052 & .049 & .143 & .106 & .045 & .040 & .151 & .114 & .042 & .039 \\ 
&      & 1000 & .086 & .080 & .025 & .031 & .094 & .084 & .021 & .026 & .097 & .093 & .022 & .026 \\ 
& .75 & 250 & .188 & .108 & .042 & .031 & .192 & .112 & .030 & .025 & .203 & .120 & .022 & .023 \\ 
&      & 500 & .122 & .089 & .024 & .017 & .130 & .092 & .014 & .014 & .135 & .099 & .015 & .013 \\ 
&      & 1000 & .090 & .067 & .015 & .009 & .091 & .069 & .008 & .008 & .093 & .075 & .006 & .008 \\ 
1.0 & .25 & 250 & .266 & .119 & .358 & .429 & .309 & .124 & .365 & .314 & .344 & .128 & .374 & .280 \\ 
&      & 500 & .186 & .124 & .294 & .367 & .251 & .128 & .291 & .262 & .281 & .133 & .291 & .238 \\ 
&      & 1000 & .137 & .126 & .219 & .315 & .174 & .130 & .223 & .231 & .205 & .137 & .224 & .216 \\ 
& .50 & 250 & .272 & .195 & .184 & .195 & .296 & .202 & .185 & .153 & .327 & .216 & .184 & .148 \\ 
&      & 500 & .169 & .182 & .098 & .124 & .192 & .189 & .092 & .095 & .217 & .204 & .098 & .093 \\ 
&      & 1000 & .102 & .161 & .060 & .075 & .114 & .169 & .051 & .060 & .124 & .185 & .051 & .062 \\ 
& .75 & 250 & .210 & .193 & .056 & .071 & .231 & .201 & .052 & .057 & .238 & .220 & .050 & .054 \\ 
&      & 500 & .139 & .161 & .025 & .037 & .157 & .169 & .032 & .030 & .155 & .188 & .025 & .028 \\ 
&      & 1000 & .093 & .127 & .013 & .019 & .101 & .134 & .016 & .017 & .102 & .151 & .011 & .016 \\ 
2.0 & .25 & 250 & .306 & .140 & .519 & .607 & .343 & .142 & .509 & .430 & .396 & .142 & .516 & .356 \\ 
&      & 500 & .304 & .161 & .471 & .572 & .352 & .163 & .477 & .396 & .355 & .164 & .498 & .330 \\ 
&      & 1000 & .241 & .174 & .410 & .539 & .292 & .176 & .430 & .380 & .301 & .179 & .464 & .325 \\ 
& .50 & 250 & .355 & .297 & .322 & .413 & .378 & .302 & .327 & .302 & .402 & .309 & .347 & .271 \\ 
&      & 500 & .277 & .297 & .207 & .301 & .306 & .303 & .213 & .218 & .338 & .314 & .234 & .202 \\ 
&      & 1000 & .175 & .283 & .117 & .201 & .191 & .290 & .120 & .151 & .219 & .305 & .125 & .149 \\ 
& .75 & 250 & .261 & .354 & .106 & .169 & .294 & .364 & .102 & .132 & .317 & .384 & .112 & .128 \\ 
&      & 500 & .180 & .317 & .052 & .088 & .211 & .328 & .048 & .069 & .223 & .350 & .052 & .066 \\ 
&      & 1000 & .123 & .270 & .028 & .043 & .143 & .281 & .031 & .036 & .146 & .305 & .026 & .036 \\
	\end{tabular} \caption{RMSE for parameters in DGP2 \eqref{eq:DGP2} for different specifications. The estimators arranged in columns are the ML estimator for $d$ ($\hat{d}^{M\!L}$), the exact local Whittle estimator for $d$ ($\hat{d}^{E\!W}$), the ML estimator for the cointegration space ($\vartheta^{M\!L}$) and narrow band least squares for the cointegration space ($\vartheta^{N\!B}$). The RMSE for cointegration spaces is based on the sine of the angle $\vartheta$ between the true and the estimated space \eqref{eq:angle}.} \label{tab:dgp2_rmse}
\end{table}

\begin{table}[ht]
	\hspace{-1.3cm}
	\begin{tabular}{rrr|rrrr|rrrr|rrrr}
		&& & \multicolumn{4}{c|}{$e=0$} & \multicolumn{4}{c|}{$e=0.5$} & \multicolumn{4}{c}{$e=1$} \\
		c & d & n & $\hat{d}^{M\!L}$ & $\hat{d}^{E\!W}$ & $\beta^{M\!L}$ & $\beta^{N\!B}$ & $\hat{d}^{M\!L}$ & $\hat{d}^{E\!W}$ & $\beta^{M\!L}$ & $\beta^{N\!B}$ & $\hat{d}^{M\!L}$ & $\hat{d}^{E\!W}$ & $\beta^{M\!L}$ & $\beta^{N\!B}$ \\
		\hline
.5 & .25 & 250 & -.017 & -.046 & -.023 & -.239 & .006 & -.047 & -.038 & -.179 & .019 & -.049 & .000 & -.198 \\ 
&      & 500 & -.006 & -.053 & -.008 & -.188 & .012 & -.057 & -.012 & -.146 & .023 & -.063 & -.003 & -.168 \\ 
&      & 1000 & -.004 & -.055 & -.006 & -.156 & .010 & -.058 & -.011 & -.123 & .021 & -.066 & -.001 & -.143 \\ 
& .50 & 250 & .027 & -.069 & .002 & -.071 & .035 & -.073 & .000 & -.054 & .032 & -.083 & .002 & -.065 \\ 
&      & 500 & .018 & -.065 & .004 & -.041 & .030 & -.067 & .002 & -.031 & .036 & -.079 & .001 & -.037 \\ 
&      & 1000 & .029 & -.057 & -.001 & -.023 & .043 & -.060 & .000 & -.019 & .042 & -.071 & .001 & -.022 \\ 
& .75 & 250 & .046 & -.061 & .001 & -.015 & .046 & -.066 & .001 & -.011 & .043 & -.076 & .000 & -.014 \\ 
&      & 500 & .035 & -.057 & .002 & -.006 & .037 & -.061 & .001 & -.006 & .037 & -.068 & -.000 & -.005 \\ 
&      & 1000 & .059 & -.045 & .001 & -.003 & .057 & -.047 & .000 & -.002 & .053 & -.055 & .000 & -.003 \\ 
1.0 & .25 & 250 & -.035 & -.092 & -.102 & -.556 & -.016 & -.091 & -.052 & -.383 & -.006 & -.089 & -.041 & -.377 \\ 
&      & 500 & -.017 & -.110 & -.023 & -.473 & -.002 & -.111 & .012 & -.344 & -.002 & -.116 & .004 & -.342 \\ 
&      & 1000 & -.008 & -.117 & -.017 & -.422 & .002 & -.121 & -.007 & -.308 & .002 & -.130 & -.016 & -.319 \\ 
& .50 & 250 & .014 & -.173 & -.006 & -.213 & .014 & -.178 & .002 & -.166 & .027 & -.191 & .000 & -.185 \\ 
&      & 500 & .007 & -.166 & .007 & -.136 & .007 & -.172 & .009 & -.106 & .012 & -.187 & .006 & -.121 \\ 
&      & 1000 & .013 & -.151 & -.002 & -.083 & .026 & -.158 & .000 & -.067 & .023 & -.175 & .001 & -.078 \\ 
& .75 & 250 & .029 & -.152 & .000 & -.051 & .026 & -.159 & .002 & -.039 & .023 & -.183 & .001 & -.047 \\ 
&      & 500 & .022 & -.135 & .003 & -.022 & .024 & -.140 & .003 & -.019 & .017 & -.160 & -.000 & -.022 \\ 
&      & 1000 & .037 & -.113 & .000 & -.010 & .041 & -.117 & .000 & -.008 & .037 & -.133 & .000 & -.010 \\ 
2.0 & .25 & 250 & -.054 & -.119 & -.367 & -.829 & -.023 & -.117 & -.275 & -.555 & -.046 & -.111 & -.241 & -.487 \\ 
&      & 500 & -.037 & -.148 & -.156 & -.766 & -.014 & -.148 & -.144 & -.536 & -.059 & -.148 & -.278 & -.470 \\ 
&      & 1000 & -.020 & -.167 & -.077 & -.731 & -.005 & -.169 & -.084 & -.517 & -.037 & -.172 & -.253 & -.470 \\ 
& .50 & 250 & -.043 & -.286 & -.053 & -.525 & -.048 & -.285 & -.028 & -.369 & -.047 & -.290 & -.008 & -.358 \\ 
&      & 500 & -.034 & -.291 & -.009 & -.375 & -.036 & -.294 & .004 & -.276 & -.023 & -.307 & .003 & -.284 \\ 
&      & 1000 & -.013 & -.278 & -.008 & -.261 & -.001 & -.285 & .005 & -.205 & .001 & -.300 & .004 & -.215 \\ 
& .75 & 250 & -.033 & -.347 & -.007 & -.168 & -.041 & -.354 & -.001 & -.127 & -.020 & -.372 & .004 & -.148 \\ 
&      & 500 & -.035 & -.310 & .002 & -.078 & -.033 & -.320 & .002 & -.062 & -.021 & -.339 & .000 & -.071 \\ 
&      & 1000 & -.000 & -.265 & -.002 & -.032 & -.005 & -.277 & -.001 & -.027 & .001 & -.299 & .001 & -.033 \\ 
	\end{tabular} \caption{Median errors for parameters in DGP2 \eqref{eq:DGP2} for different specifications. The estimators arranged in columns are the ML estimator for $d$ ($\hat{d}^{M\!L}$), the exact local Whittle estimator for $d$ ($\hat{d}^{E\!W}$), the ML estimator for the cointegration coefficient ($\beta^{M\!L}$) and narrow band least squares for the cointegration coefficient ($\beta^{N\!B}$).} \label{tab:dgp2_mede}
\end{table}


\begin{table}[ht]
	\centering
	\vspace{-0.6cm}
	\begin{tabular}{rrrr|rrrrrrr}
		a & d2 & d1 & n & $\hat{d}_1^{M\!L}$ & $\hat{d}_1^{E\!W}$ & $\hat{d}_2^{M\!L}$ & $\hat{d}_2^{E\!W}$ & $\vartheta_1^{M\!L}$ & $\vartheta_1^{N\!B}$ & $\hat{r}^{M\!L}$ \\
		\hline
		.5 & .2 & .6 & 250 & .154 & .142 & .306 & .172 & .260 & .070 & .121 \\ 
		&  &  & 500 & .176 & .112 & .310 & .143 & .326 & .051 & .090 \\ 
		&  &  & 1000 & .107 & .081 & .207 & .126 & .211 & .038 & .059 \\ 
		&  & .8 & 250 & .148 & .133 & .268 & .172 & .177 & .035 & .059 \\ 
		&  &  & 500 & .165 & .106 & .264 & .140 & .215 & .023 & .034 \\ 
		&  &  & 1000 & .121 & .079 & .166 & .123 & .156 & .014 & .014 \\ 
		& .4 & .6 & 250 & .156 & .137 & .249 & .214 & .400 & .122 & .177 \\ 
		&  &  & 500 & .136 & .108 & .201 & .178 & .458 & .102 & .162 \\ 
		&  &  & 1000 & .109 & .079 & .167 & .150 & .461 & .085 & .145 \\ 
		&  & .8 & 250 & .199 & .131 & .302 & .214 & .363 & .064 & .122 \\ 
		&  &  & 500 & .235 & .105 & .290 & .176 & .470 & .048 & .104 \\ 
		&  &  & 1000 & .279 & .079 & .306 & .147 & .581 & .036 & .075 \\ 
		2.0 & .2 & .6 & 250 & .120 & .247 & .068 & .119 & .215 & .361 & .135 \\ 
		&  &  & 500 & .082 & .216 & .045 & .089 & .151 & .222 & .117 \\ 
		&  &  & 1000 & .058 & .182 & .030 & .066 & .093 & .124 & .062 \\ 
		&  & .8 & 250 & .103 & .248 & .064 & .113 & .095 & .137 & .062 \\ 
		&  &  & 500 & .071 & .200 & .041 & .086 & .059 & .074 & .032 \\ 
		&  &  & 1000 & .052 & .160 & .028 & .066 & .033 & .045 & .011 \\ 
		& .4 & .6 & 250 & .122 & .180 & .068 & .127 & .435 & .756 & .224 \\ 
		&  &  & 500 & .084 & .164 & .048 & .107 & .369 & .675 & .213 \\ 
		&  &  & 1000 & .061 & .147 & .034 & .085 & .288 & .551 & .192 \\ 
		&  & .8 & 250 & .110 & .226 & .067 & .124 & .214 & .331 & .172 \\ 
		&  &  & 500 & .076 & .193 & .045 & .092 & .153 & .205 & .149 \\ 
		&  &  & 1000 & .055 & .164 & .031 & .069 & .096 & .129 & .126 \\ 
	\end{tabular} \caption{RMSE for parameters in DGP3 \eqref{eq:DGP3} with $r=0.5$. The estimators arranged in columns are the ML estimators for $d_1$ and $d_2$ ($\hat{d}_1^{M\!L}$ and $\hat{d}_2^{M\!L}$), the EW estimator for $d_1$ and $d_2$ ($\hat{d}_1^{E\!W}$ and $\hat{d}_2^{E\!W}$), the ML and NBLS estimators for the cointegration space ${\cal S}^{(1)}$ ($\vartheta_1^{M\!L}$ and $\vartheta_1^{N\!B}$), as well as ML for $r$ ($\hat{r}^{M\!L}$). The RMSE for cointegration spaces is based on the sine of the angle $\vartheta_j$ between the true and the estimated space \eqref{eq:angle}.} \label{tab:dgp3_r05}
\end{table}

%

\begin{table}[ht]
	\centering
	\vspace{-0.6cm}
	\begin{tabular}{rrrr|rrrrrrrr}
		a & d2 & d1 & n &  $\hat{d}_1^{M\!L}$ & $\hat{d}_2^{M\!L}$ & $\vartheta_1^{M\!L}$ & $\vartheta_1^{R\!M\!L}$ & $\vartheta_1^{N\!B}$ & $\vartheta_2^{R\!M\!L}$ & $\vartheta_2^{N\!B}$ & $\hat{r}^{M\!L}$ \\
		\hline
		.5 & .2 & .6 & 250 & .149 & .160 & .170 & .158 & .095 & .096 & .093 & .096 \\ 
		&  &  & 500 & .132 & .139 & .151 & .085 & .080 & .054 & .057 & .054 \\ 
		&  &  & 1000 & .091 & .082 & .051 & .042 & .065 & .029 & .034 & .029 \\ 
		&  & .8 & 250 & .145 & .134 & .075 & .047 & .039 & .033 & .038 & .033 \\ 
		&  &  & 500 & .099 & .089 & .048 & .018 & .028 & .012 & .018 & .012 \\ 
		&  &  & 1000 & .067 & .059 & .032 & .008 & .019 & .005 & .009 & .005 \\ 
		& .4 & .6 & 250 & .179 & .259 & .319 & .324 & .209 & .148 & .133 & .148 \\ 
		&  &  & 500 & .157 & .176 & .329 & .239 & .196 & .104 & .099 & .104 \\ 
		&  &  & 1000 & .176 & .197 & .573 & .135 & .180 & .070 & .072 & .070 \\ 
		&  & .8 & 250 & .184 & .202 & .251 & .101 & .092 & .064 & .049 & .064 \\ 
		&  &  & 500 & .185 & .202 & .340 & .050 & .074 & .034 & .025 & .034 \\ 
		&  &  & 1000 & .120 & .121 & .216 & .023 & .059 & .016 & .015 & .016 \\ 
		2.0 & .2 & .6 & 250 & .126 & .079 & .211 & .205 & .203 & .064 & .120 & .064 \\ 
		&  &  & 500 & .090 & .052 & .126 & .121 & .206 & .033 & .073 & .033 \\ 
		&  &  & 1000 & .062 & .033 & .064 & .061 & .196 & .014 & .029 & .014 \\ 
		&  & .8 & 250 & .095 & .060 & .056 & .054 & .094 & .012 & .040 & .012 \\ 
		&  &  & 500 & .061 & .042 & .022 & .020 & .077 & .003 & .014 & .003 \\ 
		&  &  & 1000 & .042 & .030 & .010 & .010 & .060 & .002 & .005 & .002 \\ 
		& .4 & .6 & 250 & .128 & .082 & .379 & .373 & .820 & .092 & .276 & .092 \\ 
		&  &  & 500 & .091 & .057 & .312 & .290 & .542 & .068 & .247 & .068 \\ 
		&  &  & 1000 & .075 & .040 & .231 & .213 & .362 & .044 & .189 & .044 \\ 
		&  & .8 & 250 & .120 & .067 & .165 & .161 & .210 & .050 & .094 & .050 \\ 
		&  &  & 500 & .077 & .047 & .080 & .074 & .204 & .018 & .040 & .018 \\ 
		&  &  & 1000 & .049 & .034 & .031 & .029 & .183 & .005 & .014 & .005 \\ 
	\end{tabular} \caption{RMSE for parameters in DGP3 \eqref{eq:DGP3} with $r=1$. The estimators arranged in columns are the ML estimator for $d_1$ and $d_2$ ($\hat{d}_1^{M\!L}$ and $\hat{d}_2^{M\!L}$), the restricted ML (setting $r=1$), the ML and NBLS estimator for the cointegration space ${\cal S}^{(1)}$ ($\vartheta_1^{R\!M\!L}$, $\vartheta_1^{M\!L}$ and $\vartheta_1^{N\!B}$), the restricted ML (setting $r=1$) and NBLS estimator for the cointegration subspace ${\cal S}^{(2)}$ ($\vartheta_2^{R\!M\!L}$ and $\vartheta_2^{N\!B}$), as well as ML for $r$ ($\hat{r}^{M\!L}$). The RMSE for cointegration spaces is based on the sine of the angle $\vartheta_j$ between the true and the estimated space \eqref{eq:angle}.} \label{tab:dgp3_r1}
\end{table}


\begin{table}[ht]
	\centering
	\vspace{-0.6cm}
	\begin{tabular}{rrrr|rrrrrrrr}
		d2 & d1 & p & n & $\hat{d}_1^{M\!L}$ & $\hat{d}_1^{E\!W}$ & $\hat{d}_2^{M\!L}$ & $\hat{d}_2^{E\!W}$  & $\vartheta_1^{M\!L}$ &$\vartheta_1^{N\!B}$ & $\vartheta_2^{M\!L}$ &$\vartheta_2^{N\!B}$ \\
		\hline
		.2 & .6 & 3 & 250 & .125 & .263 & .108 & .158 & .074 & .107 & .037 & .057 \\ 
		&  &  & 500 & .087 & .224 & .078 & .131 & .046 & .075 & .020 & .035 \\ 
		&  &  & 1000 & .064 & .187 & .055 & .115 & .032 & .048 & .013 & .023 \\ 
		&  & 10 & 250 & .063 & .268 & .069 & .157 & .013 & .029 & .013 & .036 \\ 
		&  &  & 500 & .044 & .226 & .046 & .129 & .008 & .019 & .009 & .033 \\ 
		&  &  & 1000 & .029 & .191 & .030 & .115 & .006 & .013 & .006 & .027 \\ 
		&  & 50 & 250 & .054 & .271 & .059 & .150 & .002 & .006 & .002 & .007 \\ 
		&  &  & 500 & .037 & .228 & .039 & .128 & .001 & .004 & .001 & .006 \\ 
		&  &  & 1000 & .028 & .189 & .026 & .113 & .001 & .002 & .001 & .005 \\ 
		& .8 & 3 & 250 & .104 & .274 & .108 & .167 & .033 & .047 & .018 & .028 \\ 
		&  &  & 500 & .077 & .219 & .076 & .138 & .020 & .028 & .009 & .016 \\ 
		&  &  & 1000 & .059 & .171 & .053 & .119 & .013 & .017 & .005 & .009 \\ 
		&  & 10 & 250 & .064 & .281 & .070 & .164 & .007 & .013 & .012 & .036 \\ 
		&  &  & 500 & .045 & .222 & .047 & .135 & .004 & .007 & .009 & .033 \\ 
		&  &  & 1000 & .030 & .175 & .030 & .120 & .002 & .004 & .006 & .027 \\ 
		&  & 50 & 250 & .055 & .285 & .059 & .158 & .001 & .002 & .002 & .007 \\ 
		&  &  & 500 & .037 & .224 & .039 & .136 & .001 & .001 & .001 & .006 \\ 
		&  &  & 1000 & .029 & .173 & .027 & .118 & .000 & .001 & .001 & .005 \\ 
		.4 & .6 & 3 & 250 & .121 & .237 & .121 & .190 & .163 & .192 & .034 & .059 \\ 
		&  &  & 500 & .078 & .204 & .081 & .150 & .124 & .159 & .020 & .035 \\ 
		&  &  & 1000 & .055 & .173 & .054 & .125 & .091 & .129 & .014 & .024 \\ 
		&  & 10 & 250 & .062 & .241 & .065 & .188 & .028 & .051 & .011 & .024 \\ 
		&  &  & 500 & .044 & .204 & .045 & .149 & .017 & .043 & .008 & .017 \\ 
		&  &  & 1000 & .029 & .175 & .030 & .124 & .013 & .035 & .005 & .011 \\ 
		&  & 50 & 250 & .054 & .243 & .059 & .184 & .004 & .010 & .002 & .004 \\ 
		&  &  & 500 & .036 & .207 & .038 & .148 & .003 & .008 & .001 & .003 \\ 
		&  &  & 1000 & .028 & .174 & .027 & .123 & .002 & .006 & .001 & .002 \\ 
	\end{tabular} \caption{RMSE for parameters in DGP4 \eqref{eq:DGP4} with $a=0.5$. The estimators arranged in columns are the ML estimator for $d_1$ and $d_2$ ($\hat{d}_1^{M\!L}$ and $\hat{d}_2^{M\!L}$), the EW estimator for $d_1$ and $d_2$ ($\hat{d}_1^{E\!W}$ and $\hat{d}_2^{E\!W}$), the ML and NBLS estimator for the cointegration space ${\cal S}^{(1)}$ ($\vartheta_1^{M\!L}$ and $\vartheta_1^{N\!B}$), and the ML and NBLS estimator for the cointegration subspace ${\cal S}^{(2)}$ ($\vartheta_2^{M\!L}$ and $\vartheta_2^{N\!B}$). The RMSE for cointegration spaces is based on the sine of the angle $\vartheta_j$ between the true and the estimated space \eqref{eq:angle}.} \label{tab:dgp4}
\end{table}

%% file: SecA3_Figures.tex
\begin{figure}[t!]
\begin{center}\vspace{-0.5 cm}\vspace{-0.5cm}
\includegraphics[width=\textwidth]{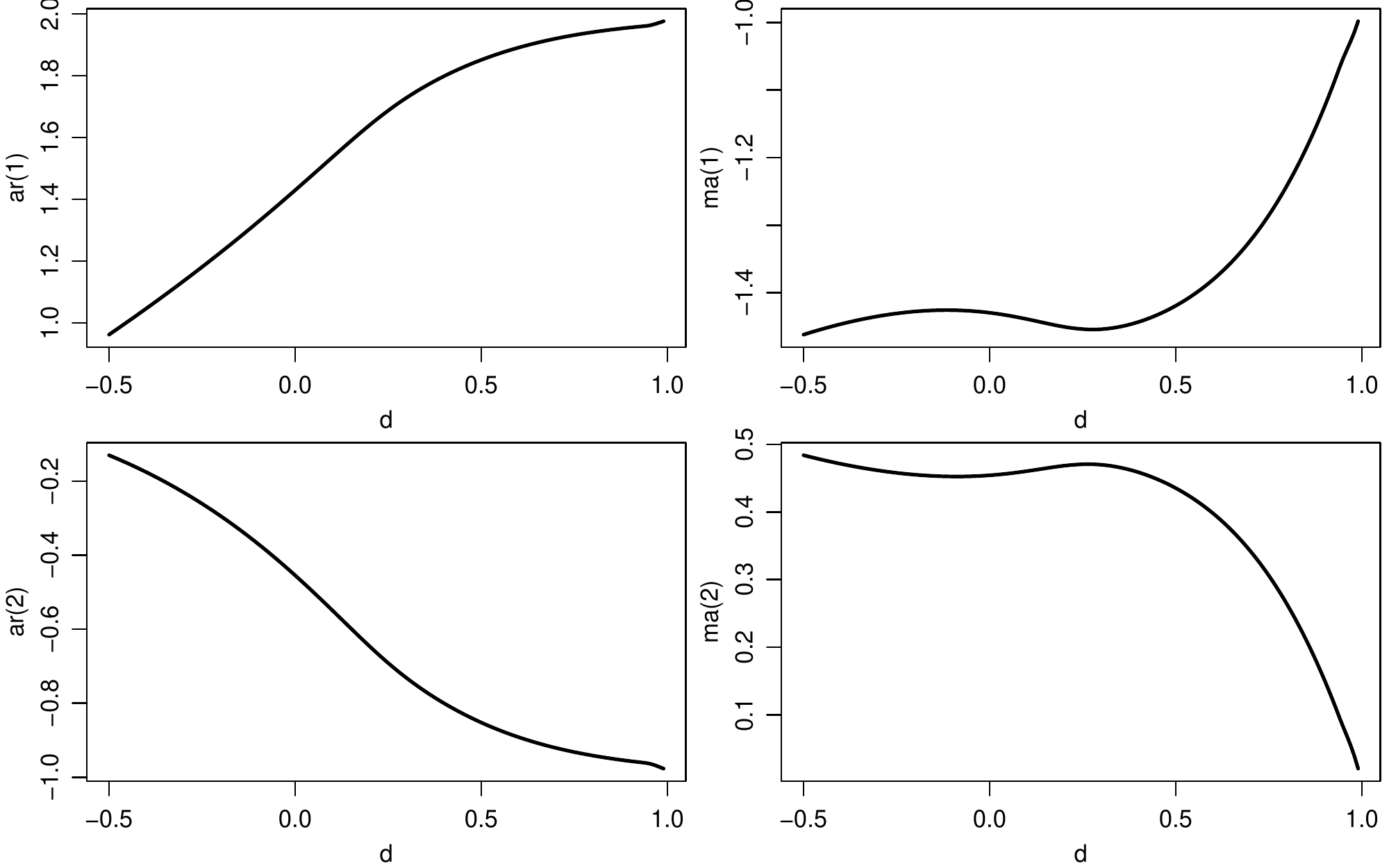}
\end{center}\vspace{-1cm} \caption{ARMA(2,2) coefficients \eqref{eq:arma_2} in the approximation of fractional processes for $d \in [-0.5 ; 1]$ and $n=500$.} \label{fig:d2arma}
\end{figure}
\begin{figure}[b!]
	\begin{center}\vspace{-0.5 cm}
		\includegraphics[width=\textwidth]{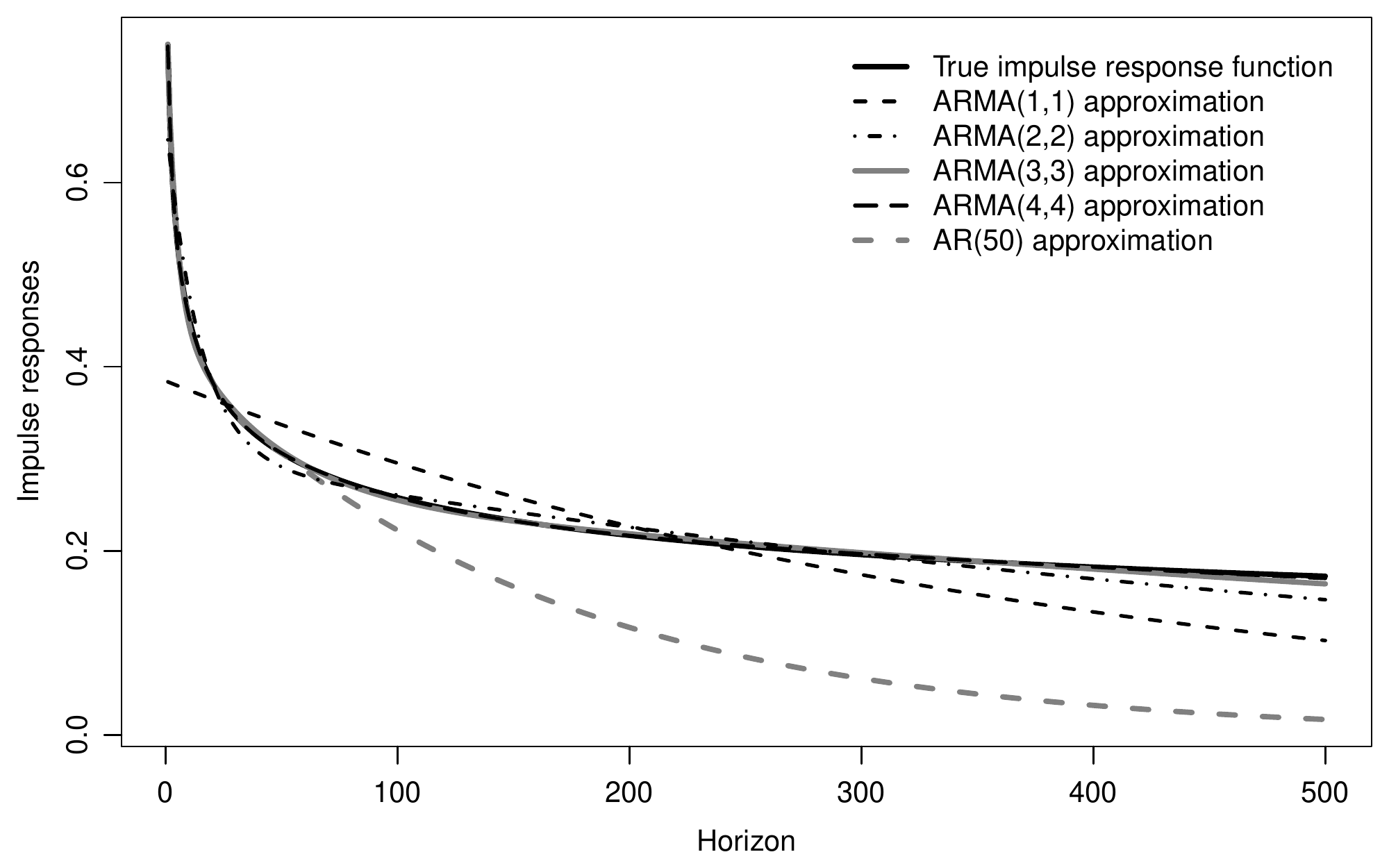}
	\end{center}\vspace{-1cm} \caption{Impulse response functions $\tilde{\psi}_j$ (see \eqref{eq:MSE_d}) for different approximating models for $d=0.75$ and $n=500$.} \label{fig:irarma}
\end{figure}
\begin{figure}[t!]
	\begin{center}\vspace{-0.5 cm}
		\includegraphics[width=\textwidth]{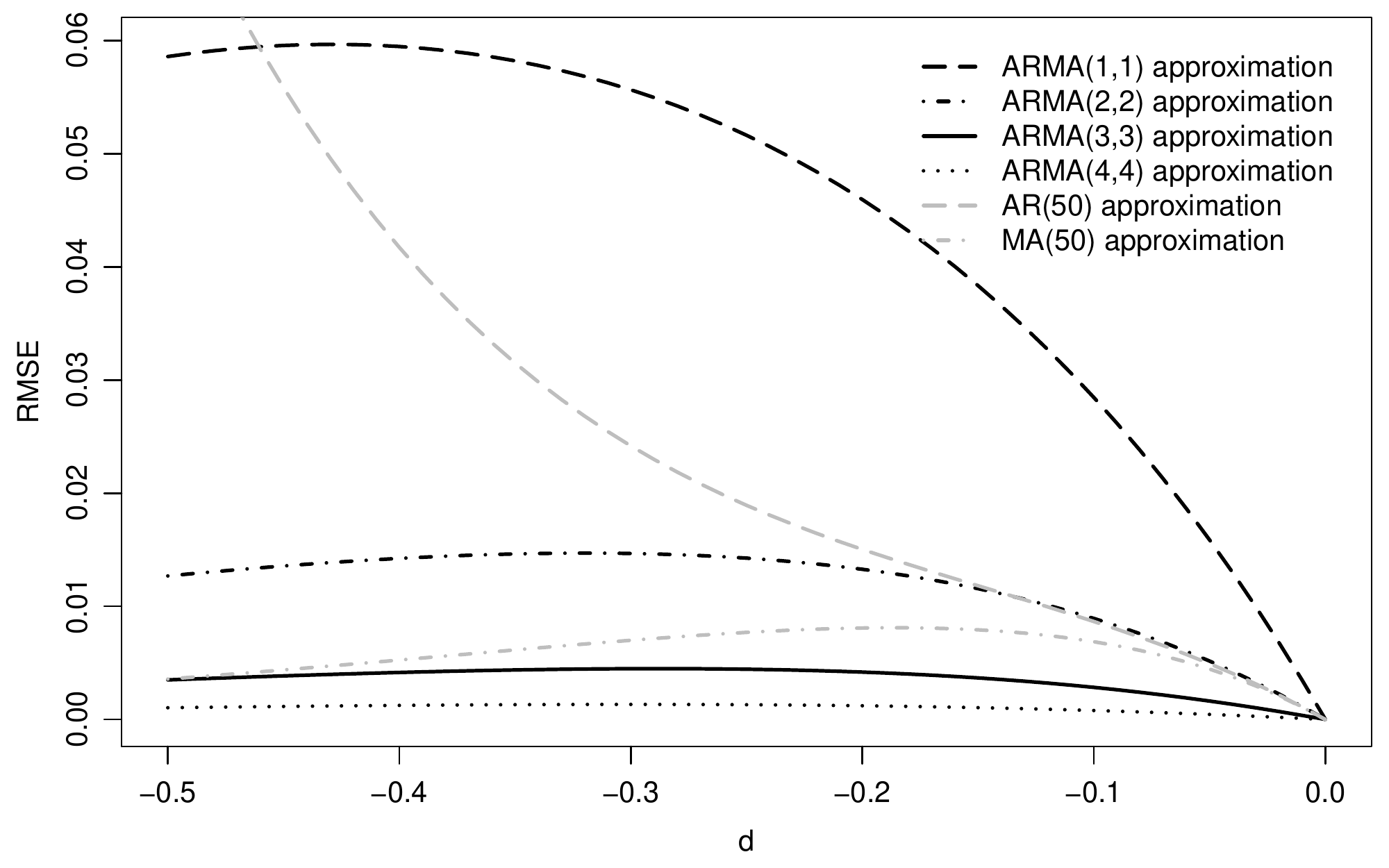}
	\end{center}\vspace{-1cm} \caption{Root mean squared error (square root of \eqref{eq:MSE_d}) for different approximating models, $d \in [-0.5 ; 0]$ and $n=500$.}\label{fig:rmse1arma}
\end{figure}
\begin{figure}[b!]
	\begin{center}\vspace{-0.5 cm}
		\includegraphics[width=\textwidth]{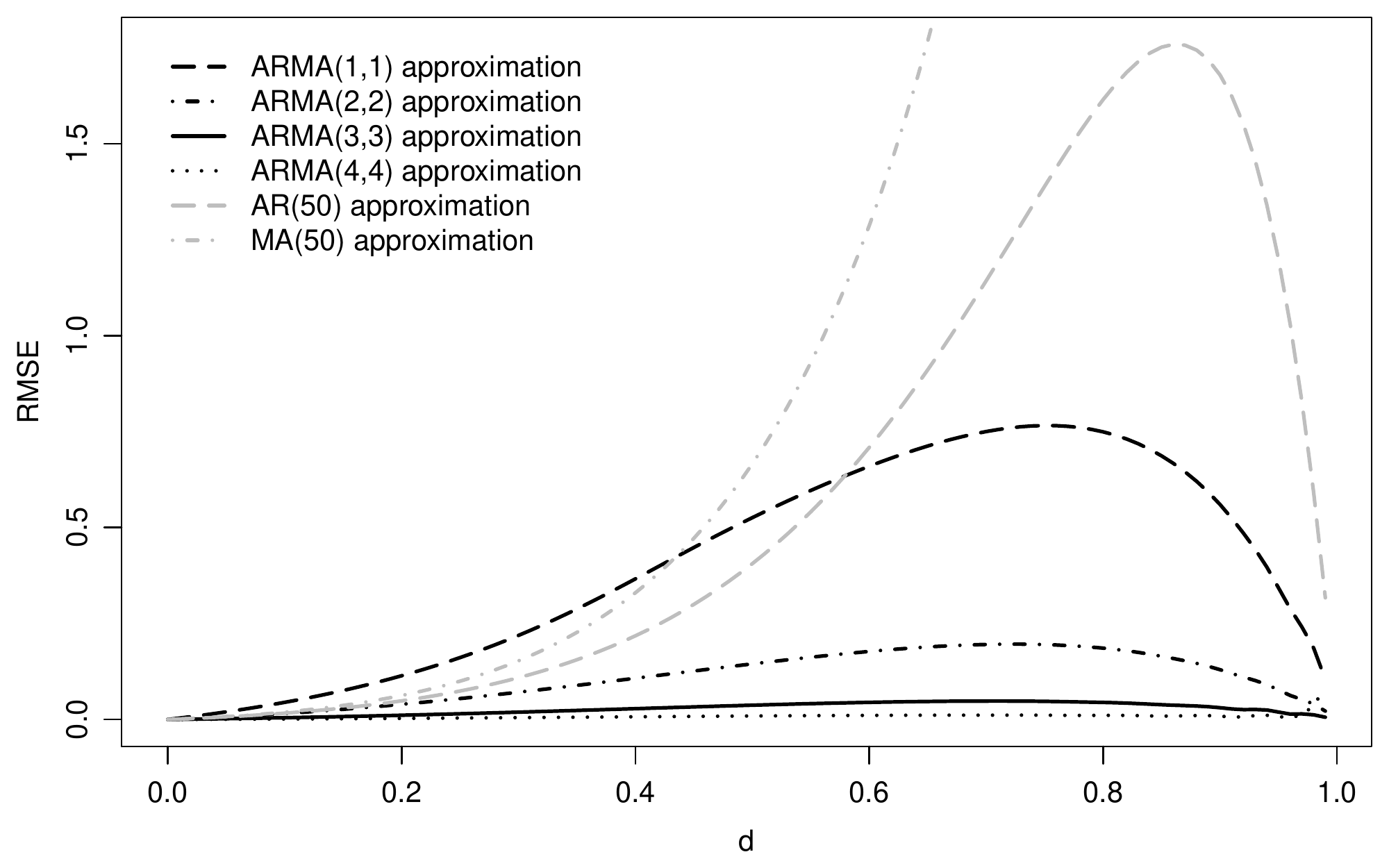}
	\end{center}\vspace{-1cm} \caption{Root mean squared error (square root of \eqref{eq:MSE_d}) for different approximating models, $d \in [0;1]$ and $n=500$.}\label{fig:rmse2arma}
\end{figure}